\documentclass[preprintnumbers,twocolumn,amsmath,amssymb,nofootinbib,prd]{revtex4-2}
\usepackage{amsmath}
\usepackage{afterpage}
\usepackage{subfigure}
\usepackage{rotating}
\usepackage{booktabs}
\usepackage{graphicx}
\usepackage{dcolumn}
\usepackage{bm}
\usepackage{float}
\usepackage{xcolor}
\usepackage{extarrows}
\usepackage{diagbox} 
\usepackage{hyperref}
\renewcommand{\arraystretch}{1.5}

\newcommand{\beq}{\begin{equation}}
\newcommand{\eeq}{\end{equation}}
\newcommand{\bea}{\begin{eqnarray}}
\newcommand{\eea}{\end{eqnarray}}
\newcommand{\bseq}{\begin{subequations}}
\newcommand{\eseq}{\end{subequations}}

\newcommand{\hf}{\frac{1}{2}}
\newcommand{\al}{\alpha}
\newcommand{\be}{\beta}
\newcommand{\ga}{\gamma}

\newcommand{\si}{\sigma}
\newcommand{\mn}{{\mu\nu}}
\newcommand{\nn}{\nonumber\\}
\def\prt{\partial}
\newcommand{\ie}{{\it i.e.}}

\newcommand{\eg}{{\it e.g.}}

\newcommand{\bit}{\begin{itemize}}
\newcommand{\eit}{\end{itemize}}

\hypersetup{
    colorlinks = true,
    allcolors = blue,
}

\providecommand{\Journal}[4] {#1 {\bf#2}, #4 (#3)}
\providecommand{\CQG}{Class. Quantum Grav.}%
\providecommand{\CPC}{Chin. Phys. C}%
\providecommand{\GRG}{Gen. Rel. Grav.}%
\providecommand{\LRR}{Living Rev. Relativ.}%
\providecommand{\PR}{Phys. Rev.} %
\providecommand{\PRep}{Phys. Rep.} %
\providecommand{\PRL}{Phys. Rev. Lett.} %
\providecommand{\PRD}{Phys. Rev. D} %
\providecommand{\PRSLA}{Proc. R. Soc. Lond. A} %
\providecommand{\RMP}{Rev. Mod. Phys.} %
\providecommand{\PLA}{Phys. Lett. A} %
\providecommand{\PLB}{Phys. Lett. B} %
\providecommand{\NPB}{Nucl. Phys. B} %

\providecommand{\EPJC}{Eur. Phys. J. C.} %
\providecommand{\APny}{Ann. Phys.(N.Y.)} %
\providecommand{\APJ}{ApJ} %
\providecommand{\JMP}{J. Math. Phys.} %
\providecommand{\JPAMT}{J. Phys. A: Math. Theor.} %
\providecommand{\JHEP}{JHEP} %
\providecommand{\JCAP}{JCAP} %
\providecommand{\IJMPD}{Int. J. Mod. Phys. D}%
\begin{document}

\title{\Large {Maxwell theories along the light track:
Null Formalism in extended electrodynamics}}
\author{Zhi Xiao${}^{a, b}$}
\email{spacecraft@pku.edu.cn}
\affiliation{${}^{a}$North China Electric Power University, Beijing 102206, China}
\affiliation{${}^{b}$Hebei Key Laboratory of Physics and Energy Technology, North China Electric Power University, Baoding 071000, China}

\author{Bing Sun${}^{c}$}
\email{bingsun@bua.edu.cn}
\affiliation{${}^{c}$Department of Basic Courses, Beijing University of Agriculture, Beijing 102206, China}

\author{Tao Zhu${}^{d, e}$}
\email{zhu05@zjut.edu.cn}
\affiliation{${}^{c}$Institute for Theoretical Physics \& Cosmology, Zhejiang University of Technology, Hangzhou, 310023, China}
\affiliation{${}^{d}$United Center for Gravitational Wave Physics (UCGWP), Zhejiang University of Technology, Hangzhou, 310023, China}

\begin{abstract}
We develop a differential-form approach to systematically derive the Newman-Penrose null-tetrad equations for Lorentz-violating extensions of Maxwell electrodynamics. 
The coordinate-independent nature of differential forms allows the actions and corresponding field equations of the theory to be expressed compactly and enables a systematic and transparent derivation of first-order equations in the Newman-Penrose formalism.
Within this formalism, we explicitly present a simple algebraic construction for the gauge invariant extended Maxwell actions that avoids explicit index manipulations up to mass dimension six. The combined scheme of differential-form approach and Newman-Penrose formalism offers an efficient tool for analyzing Lorentz-violating effects on asymptotic photon propagation and polarization.
\end{abstract}
\maketitle

\section{Introduction}
It is remarkable that both gauge invariance and Lorentz symmetry are intrinsically encoded in Maxwell’s electrodynamics.
The non-abelian generalization of the former underlies the basic structure of the particle Standard Model (SM),
while the latter is embedded in the foundations of both the SM and general relativity. 
However, decades of unsuccessful attempts to construct a
satisfactory theory of quantum gravity
suggest the possibility that Lorentz symmetry may be violated,
as indicated in frameworks such as bosonic string field theory \cite{AKS89}, loop quantum gravity \cite{loop99} and noncommutative field theory \cite{nonCommF} (for a comprehensive review, see \cite{LVMotivations}).
Moreover, sustained efforts led by Kosteleck\'y and collaborators have established a comprehensive effective field theory (EFT) framework incorporating Lorentz violation (LV), the Standard-Model Extension (SME) \cite{SMEa, SMEbmini, SMEb, SMEgauge, SMEg}.
The photon sector \cite{SMEa, SMEbmini, SMEb, SMEgauge} constitutes an indispensable part; consequently, extensive experimental and theoretical efforts have been devoted to tests of Lorentz invariance in electrodynamics \cite{Potting25, dataTable}.

However, up to now, most studies of LV in electrodynamics 
have been carried out either in flat spacetime or confined to local properties of photons. 
As an EFT, one would expect that 
at least the renormalizable dimension-3 and dimension-4 photon operators may also have an impact on the asymptotic or large-scale behavior of electromagnetic fields. 
This possibility has recently been investigated in \cite{ZX2025}, motivated by an insight \cite{Yuri21Asym} 
obtained in the study of the ``t-puzzle" \cite{Yuri15tPuzzle}.
The insight reveals the necessity of studying asymptotic behaviors, since a generic non-minimal Weyl tensor coupling
with $t$-coefficients
would be incompatible with non-trivial asymptotic flatness, namely, asymptotic flatness other than Minkowskian spacetime. For related modified asymptotic behaviors, 
see also the results on compact solutions in bumblebee gravity \cite{Bumblebee25}.

The natural framework for studying the asymptotic behavior of massless fields is the Newman-Penrose (NP) formalism (also known as the null formalism) \cite{NP1962-68, Chand-MToBH}, which was initiated by Bondi and Sachs and later elaborated by Newman and Penrose.
It provides a scalar formulation of the field equations, 
effectively reducing the equations of motion from partial differential equations to first-order ordinary differential equations and thereby greatly simplifying their solution. 
Moreover, it offers several additional analytical advantages,
including a clean separation of the physical modes manifested in the celebrated peeling theorem \cite{Penrose60s, NU1962Asym},
invariant classification based on symmetries \cite{PetrovC},
perturbation analysis on a given background spacetime \cite{Teukolsky1973}, and the analysis of polarization evolution along null congruence in generic curved spacetimes \cite{Frolov20s, Dolan18}.

On the other hand, the Faraday tensor in electrodynamics,
as well as the coframe and spin-connection in gravity
admit elegant formulations in terms of differential forms (DFs), leading to concise and coordinate-free expressions of
the field equations. 
For example, the homogeneous Maxwell equation follows immediately from the definition $F=dA$, which implies $dF=0$, while the second Bianchi identity $DR=0$, 
satisfied by the curvature 2-form $R=d\omega+\omega\wedge\omega$
with $\omega$ the spin connection, 
arises as a geometric consequence 
of the Cartan structure equations and the non-abelian nature of Lorentz gauge symmetry.
Moreover, in the Lorentz-violating (LV) theories, the DF formalism can be used neatly to derive covariant dispersion relations \cite{Itin0709} without imposing gauge-fixing condition, 
for both photon and gravitational waves \cite{SMEb, SMEgauge, LVGW18}.


While certain elements of the null formalism, most notably spin-weighted spherical harmonics, 
have been extensively used to analyze LV effects in electromagnetic radiation \cite{SMEb}, 
explicit use of the null tetrad remains uncommon. 
In our previous work, we derived the CPT-odd Maxwell equations in the NP formalism 
and analyzed their asymptotic behavior in spherically symmetric spacetimes \cite{ZX2025, XWS2025}.
Here, we further develop the differential form approach and combine it with the null formalism 
to investigate more general LV extended Maxwell equations.
A key challenge is that the nullity condition $p^2=0$ (with respect to the background metric) is generally violated in LV theories. 
Nevertheless, for perturbative treatments -- provided that one avoids regimes dominated by non-perturbative effects
-- the null tetrad remains a valid leading-order approximation, given that Lorentz symmetry is experimentally well upheld \cite{dataTables}.

More broadly, null methods are widely employed in extended gravity theories, many of which incorporate LV.
Examples include the classification of multi-metric theories \cite{GWTestGR};
asymptotic and memory-effect analyses in both dynamical Chern–Simons gravity \cite{CSGMemory} 
(which preserves a restricted local Lorentz symmetry but violates parity) 
and Einstein-\AE ther theory \cite{EAEMemory};
and investigations of the quasi-normal modes of scalar and Dirac perturbations in bumblebee gravity \cite{DiracBumblebee}.
We expect that the present work on LV extensions of Maxwell theory may help pave the way for more ambitious 
investigations of LV extended gravitational theories.

The remainder of this paper is organized as follows.
In Sec. \ref{NullTetrad}, we review the basics of the null formalism,
and derive the Maxwell equations in the Newman-Penrose (NP) framework
using two complementary approaches:
the intrinsic derivative method and the differential forms approach (DFA). 
while the DFA was previously presented in Ref. \cite{ZX2025},
we provide a more detailed and explicit derivation here.
In Sec. \ref{LVMaxwNull}, we apply the null formalism to extended Maxwell theories up to dimension five.
We derive the corresponding Maxwell equations in NP form
and present a systematic procedure for constructing Maxwell actions in differential forms up to dimension six.
Finally, Sec. \ref{summ} summarizes our main results.

The conventions are the following.
Our metric signature is $(-,+,+,+)$ and the totally antisymmetric Levi-civita tensor is $\epsilon_{0123}=+1$.
The Greek indices $\mu,~\nu,~\rho,...$ are for the spacetime manifold, and the Latin indices $a,b,c,...$ are for
frame or tetrad. An overbar, such as $\bar{A}$, denotes the complex conjugate of a generic complex quantity $A$.
The bold notation is left for null 1-forms, such as ${\bf{l}}=l_\mu dx^\mu$.
The round bracket and square bracket are for symmetrization and anti-symmetrizations, respectively,
say, $A_{[a_1...a_n]}\equiv\frac{1}{n!}\sum_\mathcal{P}(-1)^{\mathrm{sign}(\mathcal{P})}A_{\mathcal{P}(a_1...a_n)}$ and $A_{(a_1...a_n)}\equiv\frac{1}{n!}\sum_\mathcal{P}A_{\mathcal{P}(a_1...a_n)}$,
where $\mathcal{P}$ is the permutation of n indices.

\section{LI Null formalism and Maxwell equations}\label{NullTetrad}

In simple terms, the null formalism is a specialized form of the tetrad formalism
in which tensor equations are projected onto a basis of null vectors,
reducing them to scalar equations.
In the Newman-Penrose notation \cite{NP1962-68}, the null tetrad consists of four vectors:
two real null vectors $l^\mu,~n^\mu$ associated with outgoing and ingoing null directions, respectively;
and a pair of complex conjugate null vectors $m^\mu,~\bar{m}^\mu$ that span the transverse spatial directions.
Often the tetrad is chosen to be adapted to the symmetries of the underlying spacetime,
which can greatly simplify the field equations
and endow quantities such as the spin coefficients with clear physical interpretations.

In the case of electromagnetism, the Faraday tensor $F_{\mn}$ acquires a remarkably compact
representation: it can be expressed in terms of three complex scalars,
$\phi_a,~a=0,1,2$, each of which carries well-defined spin and boost weights.
These scalars neatly separate different physical components of the photon fields
as ingoing and Coulomb modes, and outgoing radiation, respectively.
They thus facilitate both analytic calculations and the asymptotic analysis.

\subsection{Preliminary of null formalism}
Next, we will briefly review some preliminaries on null formalism,
which can be found in Ref. \cite{NU1962Asym}, or in an excellent review article \cite{Newman2009}.
The Lorentz invariant null formalism is based on the following facts:\\
1. {\it The metric in tangent space is quasi-orthonormal},
\bea\label{quasi-orthon}
\eta^{ab}=g^{\mn}E^a_{~\nu}E^b_{~\mu},
\eea
where $\{E^a_{~\nu},a=1,...,4\}=(l_\nu,n_\nu,m_\nu,\bar{m}_\nu)$ are null covectors and
\bea\label{FrameMetric}
\eta^{ab}=\left(
            \begin{array}{cccc}
              0 & -1 & 0 & 0 \\
              -1 & 0 & 0 & 0 \\
              0 & 0 & 0 & 1 \\
              0 & 0 & 1 & 0 \\
            \end{array}
          \right)=\eta_{ab}.\nonumber
\eea
Here we intentionally use capital $E^a_{~\nu}$ in place of $e^a_{~\nu}$ to distinguish the null tetrad from the ordinary orthonormal tetrad consisting of one timelike vector and three 3 space vectors. \\
2. {\it The null tetrad $\{E^a_{~\mu},a=1,...,4\}$ forms a complete basis for the tangent space}, satisfying
\bea\label{completeN}
g_{\mn}=\eta_{ab}E^a_{~\mu}E^b_{~\nu}=2[m_{(\mu}\bar{m}_{\nu)}-l_{(\mu}n_{\nu)}],
\eea
where $g_{\mn}$ is the spacetime metric,
$\eta^{bc}\eta_{ab}=\delta^c_a$ and the contravariant null vectors $E_a^{~\mu}=\eta_{ab}\,g^{\mn}\,E^b_{~\nu}$.
For notational convenience, we denote the set of null vectors as $\{E_a^{~\nu},a=1,...,4\}=(-n^\nu,-l^\nu,\bar{m}^\nu,m^\nu)$
such that $E_a^{~\nu}E^a_{~\mu}=\delta_\mu^\nu$ and $E_a^{~\mu}E^b_{~\mu}=\delta_a^b$.
The two key properties of the null tetrad, namely its quasi-orthonormality and completeness relations, are in direct correspondence with those of the standard orthonormal tetrad erected for a co-moving observer along a timelike world line.
The basic picture may be outlined as follows.
\bit
\item Orthonormal tetrad (timelike world line). 
Given a timelike world line $\gamma$, we may introduce a co-moving observer whose four-velocity $u^\mu$ is tangent to $\gamma$ at each point $P\in\gamma$. At point $P$, we complete $u^\mu$ to an orthonormal tetrad by choosing three spacelike orthonormal vectors $e^{\mu}_{~i}$ ($i=1,2,3$) spanning the local rest space. They satisfy $g_{\mn}e^{\mu}_{~i}e^{\nu}_{~j}=\delta_{ij}$, $u^2=-1$ and $g^{\mn}u^\mu\,e^{\nu}_{~j}=0$. We then transport the tetrad along $\gamma$, for example by parallel transport or (more physically) by Fermi–Walker transport. 
\item Null tetrad (null congruence). 
Given a null congruence of null geodesics, we may choose a wavefront function $u(x)$ as a null coordinate labeling constant-phase hypersurfaces. Its gradient defines a null covector $l_\mu=-\nabla_\mu u$. Raising the index gives the associated null vector, $l^\mu=g^{\mn}l_\nu$, 
which is tangent to the null generators. Along each null geodesic we introduce an affine parameter $r$, so that $l^\mu=\frac{dx^\mu}{dr}$. 
We then choose a second null vector $n^\mu$ satisfying the normalization $n\cdot l=-1$. 
The $2-$dimensional subspace orthogonal to both $n^\mu$ and 
$l_\mu$, denoted as $(n\wedge l)_\perp$, is purely spacelike and can be spanned by two real orthonormal spacelike vectors $\xi^\nu,~\zeta^\nu$. Combining them into a complex null vector $m^\nu\equiv\frac{1}{\sqrt{2}}[\xi^\nu+i\zeta^\nu]$ and its complex conjugate $\bar{m}^\nu$,
we obtain the standard null tetrad 
$(l_\mu,~n^\mu,~m^\mu,~\bar{m}^\mu)$.
\eit

Similar to an orthonormal tetrad, where fixing the observer’s four-velocity $u^\mu$ leaves a $SO(3)$ rotational freedom in the choice of the spatial triad $e^{\mu}_{~i}$, a null tetrad admits a larger gauge freedom associated with the Lorentz group $SO(1,~3)$ [or equivalently $SL(2,\mathbb{C})$], since it is naturally adapted to the light cone rather than to a timelike rest frame. This reflects the fact that the propagation of a massless field is governed by the causal structure of spacetime, which is preserved under Lorentz transformations.
\begin{figure}[htbp]
 \centering
 \hspace{-0.05in}
  {\includegraphics[width=0.4\textwidth]{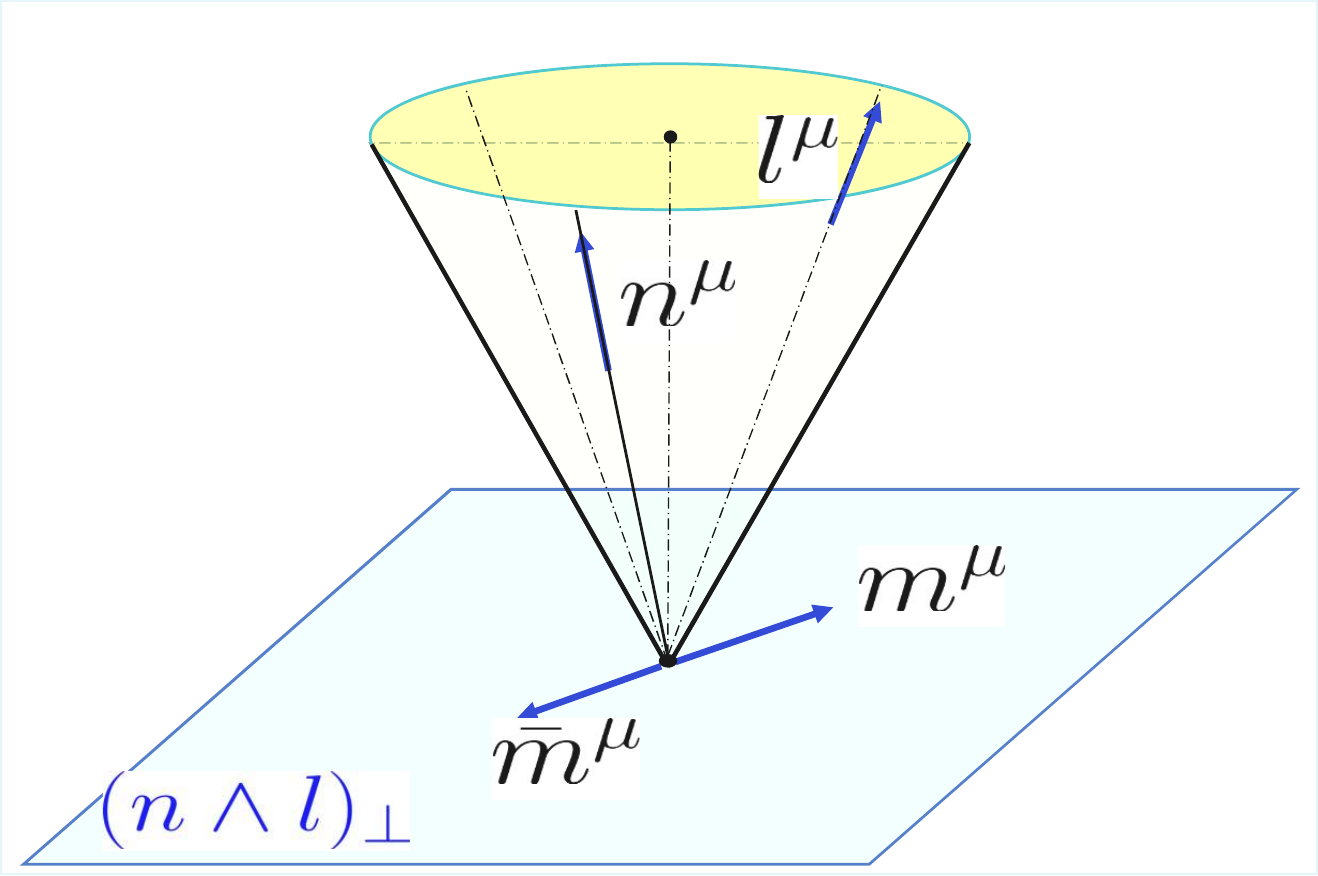}}
  \caption{Rays in a linearized radiation field. Figure is copied from the Ref. \cite{Sachs1960s}.
  In this figure, $k^a$ is denoted as the first null vector $l^\mu$, $v(x)$ is the affine parameter $r$, and $z^a[r^s(x)]$ is the source world line with four-velocity $e^a=dz^a/ds$.}\label{radiationF}
\end{figure}

For an intuitive illustration of null tetrad, 
see Fig. \ref{radiationF}.
A concrete procedure for constructing a null tetrad is given in \cite{NU1962Asym, Newman2009}. For the reader’s convenience, we also provide a brief review in appendix \ref{Procedure-NT}.


Even though in the presence of LV, massless fields need not propagate 
along null geodesics of the background metric
(namely $g_{\mn}k^\mu k^\nu=0$), 
existing LV tests indicate that Lorentz symmetry is extremely well satisfied \cite{dataTables}, 
even if it may not be exact at some level. 
Practically, we may therefore still employ the usual null tetrad as a leading-order approximation, 
provided we do not extrapolate to regimes where the perturbative treatment breaks down,
such as in the analysis of asymptotic behavior at null infinity, 
where the difference in causal structure becomes essential.

For a more systematic treatment of LV theories, 
one may instead construct a tetrad adapted to an effective metric $\tilde{g}_{\mn}(K)$,
where $K$ denotes a generic LV coefficients with indices suppressed.
Using the completeness condition (\ref{completeN}),
an effective null vector may be parameterized as
\bea\label{QuasiNull}
\tilde{l}^\mu=l^\mu(1+\rho)+\alpha\,n^\mu+\beta\,m^\mu+\bar{\beta}\,\bar{m}^\mu,
\eea
where $\rho, \alpha\in\mathbb{R},~\beta\in\mathbb{C}$ are small parameters of order
$\mathcal{O}(\epsilon)$, with $\epsilon$ the dimensionless magnitude of LV ({\it e.g.}
$\epsilon\sim\frac{\delta c}{c}$).
Then $\tilde{l}^2=-2\alpha+\mathcal{O}(\epsilon^2)$,
and analogous expansions hold for the remaining tetrad vectors, 
with the parameters constrained so that the tetrad remains quasi-orthonormal 
in the same sense as Eq. (\ref{quasi-orthon}).

However, such a LV ``null” tetrad is not universal:
it depends on the particle species and, 
more generally, 
on the specific LV theory through $\tilde{g}^{\mn}$. 
A systematic development of this framework would be interesting, 
but lies beyond the scope of the present work.
Nevertheless, as already mentioned, even the LI null tetrad has been applied to the classification of multi-metric theories \cite{GWTestGR}, to asymptotic analyses and the memory effect in modified gravity \cite{CSGMemory, EAEMemory}, and to studies of quasi-normal modes via Dirac perturbations \cite{DiracBumblebee}.
By contrast, our target here is comparatively simpler, as we aim to apply the null formalism to extended electrodynamics.

\subsection{Null projection of LI Maxwell equations}
The central task is to project Maxwell’s equations onto a null tetrad. In the null formalism, the idealized source is taken to move along a timelike world line, or, more generally, to be confined within a world tube of sufficiently compact spatial extent. In either case, the associated null geodesics -- when traced backward -- converge onto this world line (or world tube), i.e. onto the source region. Consequently, the electromagnetic field outside the world tube must satisfy the vacuum Maxwell equations.
To our knowledge, there are at least two ways \cite{ZX2025} to carry out this projection: \\
1. Use the definitions of the three complex null scalars and {\it the intrinsic derivatives to project the Maxwell equations into four first-order differential equations}.\\
2. Exploit {\it the coordinate-independence of differential forms}. Note that the null-tetrad formulation of Maxwell equations is simply the projection of the tensor-form equations onto a moving null frame,
rather than a fixed global Cartesian frame.

\subsubsection{the intrinsic derivative approach}\label{IntrinsicApproach}
The first approach to Maxwell equations can be found in the appendix of the classical paper of Newman and Penrose \cite{NP1962-68}, as well as in Chandrasekhar's textbook \cite{Chand-MToBH}.
For completeness, we provide a brief review here.

We temporarily include the charge current $j_\mu$ in the inhomogeneous Maxwell equation and collect it together with the off-shell identity as 
\bea\label{Maxwell-LI}
\nabla_\mu F^{\mu}_{~\nu}=g^{\mu\al}F_{\al\nu;\mu}=-j_\nu,\quad F_{[\mn,\rho]}=0.
\eea
The second equation is the Bianchi identity.
It may be written as $\epsilon^{\mn\rho\si}F_{\mn,\rho}=0$,
and is equivalent to the compact
differential form statement $F=dA$.
To accommodate curved coordinates, we use covariant derivatives ($\nabla_\mu$, or semicolons) instead of partial derivatives (commas), even in flat spacetime.
For clarity, we {\it take the Faraday tensor with lower indices, $F_{\mn}$, as the fundamental object}.
The upper indices tensor $F^{\mn}$, or more generally $\mathcal{H}^{\mn}$,
which in media is often referred to as the electromagnetic excitation tensor, 
is related to $F_{\mn}$ through the constitutive relation
\bea
\mathcal{H}^{\mn}=\hf\chi^{\mn\rho\si}F_{\rho\si}.
\eea
This relation applies both in material media and in vacuum. In vacuum, the constitutive pseudo-tensor takes the standard metric form
$\chi^{\mn\rho\si}=g(g^{\mu\si}g^{\nu\rho}-g^{\mn}g^{\rho\si})$, where $g^{\mn}$ is the inverse metric tensor.

For an arbitrary tensor $T^{\mu...}_{~\rho\si...}$, its components in a tetrad frame (not necessarily null) are obtained by projection onto the frame basis,
\bea
T^{a...}_{~bc...}=E^a_{~\mu}...T^{\mu...}_{~\rho\si...}E_b^{~\rho}E_c^{~\si}....\nonumber
\eea
The quantities $T^{a...}_{~bc...}$ are scalars under spacetime coordinate transformations, although they carry frame indices and thus transform under local frame rotations.
In particular, for the Faraday tensor, $F_{ab}=E_a^{~\mu}E_b^{~\nu}F_{\mn}$,
with the inverse relation $F_{\mn}=E^a_{~\mu}E^b_{~\nu}F_{ab}$.

Inspection of Eqs. (\ref{Maxwell-LI}) reveals that
the only nontrivial step in projecting them into frame-index form is handling the covariant derivatives.
At this point, the {\it intrinsic derivative} comes into play.
To our best knowledge, the concept was first introduced by \'E. Cartan \cite{ECartan83},
and was later developed extensively in the context of curved spacetime by S. Chandrasekhar \cite{Chand-MToBH}.
For clarity, we illustrate the idea using a vector as an example:
\bea\label{IntrinsicD}
A_{a|b}\equiv\,e_a^{~\mu}A_{\mu;\nu}e_b^{~\nu}\Leftrightarrow\,A_{\mu;\nu}=e^a_{~\mu}A_{a|b}e^b_{~\nu}.
\eea
Here $e_a^{~\mu}$ denotes a tetrad, 
not necessarily null.
This relation shows that the intrinsic derivative directly connects the covariant derivative of a tensor
to its frame derivative.
Expanding the right hand side of Eq. (\ref{IntrinsicD}), we get 
\bea&&
\hspace{-6mm}A_{a|b}=e_a^{~\mu}A_{\mu;\nu}e_b^{~\nu}=e_a^{~\mu}\nabla_bA_\mu=e_a^{~\mu}\nabla_\nu(e^c_{~\mu}A_{c})e_b^{~\nu}\nn&&
\hspace{-8mm}=e_a^{~\mu}[e^c_{~\mu}\prt_\nu A_{c}+\nabla_\nu e^c_{~\mu}A_c]e_b^{~\nu}
=A_{a,b}+\eta^{cd}\ga_{acb}A_d,
\eea
where $A_{c,b}=\prt_b A_c=e_b^{~\nu}\prt_\nu A_{c}$ is the directional derivative along the tetrad vector $e_b^{~\nu}$.
In the last step we introduced the spin coefficients $\ga_{acb}\equiv e_a^{~\mu}e_{c\mu;\nu}e_b^{~\nu}=-\ga_{cab}$.

Now it is time to rewrite Eqs. (\ref{Maxwell-LI}) into its frame form,
\bea&&\label{Bianchi-Maxw-f}
\hspace{-3mm}F_{[\mn;\rho]}=0\Rightarrow\epsilon^{\mn\rho\si}F_{\mn;\rho}=0\nn&&
\Rightarrow\epsilon^{abcd}F_{ab|c}=0
\Leftrightarrow\,F_{[ab|c]}=0,\\&&\label{Inhomo-Maxw-f}
\hspace{-3mm}g^{\mn}F_{\mu\rho;\nu}=-j_\rho\Rightarrow\,
\eta^{mn}F_{mb|n}E^b_{~\rho}=-j_\rho
\nn&&
\Leftrightarrow
\eta^{mn}F_{ma|n}=-j_a.
\eea
where $j_a\equiv\,E_a^{~\rho}j_\rho$ and the null tetrad are collectively denoted as
\bea&&
\{E_a^{~\nu},a=1,...,4\}=\{-n^\nu,-l^\nu,\bar{m}^\nu,m^\nu\},\nn&&
\{E^a_{~\nu},a=1,...,4\}=\{l_\nu,n_\nu,m_\nu,\bar{m}_\nu\}.
\eea
To proceed further, the six degrees of freedom of the Faraday tensor are encoded in the three complex Newman–Penrose (NP) scalars
\bea\label{NPS-Faraday}&&
\phi_0=F_{31}=F_{\mn}m^\mu l^\nu, \phi_2=F_{24}=F_{\mn}n^\mu\bar{m}^\nu, \nn&&
\phi_1=\hf[F_{21}+F_{34}]=\hf F_{\mn}[n^\mu l^\nu +m^\mu\bar{m}^\nu].
\eea
As noted earlier, these expressions can be found in Refs. \cite{NP1962-68,Chand-MToBH}. 
Equations (\ref{NPS-Faraday}) differ from the corresponding formulas in Refs. \cite{NP1962-68,Chand-MToBH} by an overall minus sign, owing to our choice of metric signature $+2$ rather than $-2$.

The remaining step is to recast Eqs. (\ref{Bianchi-Maxw-f}) and (\ref{Inhomo-Maxw-f}) in terms of the three complex NP scalars.
As an illustration, we substitute the Bianchi identity $F_{[23|4]}=0$ into the inhomogeneous Maxwell equation for $j_2$.
Starting from
\bea&&
\hspace{-3mm}-j_2=\eta^{mn}F_{m2|n}=F_{32|4}+F_{42|3}-F_{12|2}\nn&&
=F_{21|2}-F_{24|3}+F_{32|4},
\eea
and using
$F_{[23|4]}=0\Rightarrow
F_{32|4}=F_{34|2}-F_{24|3}$,
we obtain 
\bea
\hspace{-3mm}F_{24|3}-\hf[F_{21|2}+F_{34|2}]=\frac{j_2}{2}
\Leftrightarrow
\phi_{2|3}-\phi_{1|2}=\frac{j_2}{2}.
\eea
Proceeding in the same manner, substituting $F_{[34|1]}=0$ into the equation for $j_1$,
$F_{[23|1]}=0$ into that for $j_3$,
and $F_{[14|2]}=0$ into that for $j_4$,
we obtain the remaining three equations.
Collecting them together, we have
\bea\label{MaxwellNP0}&&
\phi_{1|1}-\phi_{0|4}=\frac{j_1}{2},\quad
\phi_{2|3}-\phi_{1|2}=\frac{j_2}{2},\nn&&
\phi_{1|3}-\phi_{0|2}=\frac{j_3}{2},\quad
\phi_{2|1}-\phi_{1|4}=\frac{j_4}{2}.
\eea

The final step is to express the intrinsic derivatives in terms of the directional derivatives
and spin coefficients.
As an illustration, we consider the last equation above and give the details below:
\bea&&
\hspace{-5mm}
\phi_{2|1}=
F_{24,1}+\eta^{mn}[\ga_{2m1}F_{n4}+\ga_{4m1}F_{2n}]
=\phi_{2,1}-2\pi\phi_1\nn&&
\hspace{-3mm}
-(\ga_{211}+\ga_{341})\phi_2-2\ga_{421}\phi_1
=(D+2\epsilon)\phi_2,\\&&
\hspace{-5mm}
\phi_{1|4}=\hf(F_{21|4}+F_{34|4})
=\phi_{1,4}+\hf\eta^{mn}\nn&&
\hspace{-3mm}
\cdot[\ga_{2m4}F_{n1}+\ga_{1m4}F_{2n}+\ga_{3m4}F_{n4}+\ga_{4m4}F_{3n}]\nn&&
\hspace{-3mm}
=\bar{\delta}\phi_1-(\lambda\phi_0-\rho\phi_2).
\eea
Substituting these into the corresponding Maxwell equation yields
\bea&&
(D+2\epsilon-\rho)\phi_2-(\bar{\delta}+2\pi)\phi_1+\lambda\phi_0
=\frac{j_4}{2}.
\eea
We now summarize the three remain Maxwell equations together with the upper equation in the NP formalism,
\bea&&\label{LINPMax-1form}
(D-2\rho)\phi_1-(\bar{\delta}-2\al+\pi)\phi_0+\kappa\phi_2=\frac{j_1}{2},\nn&&
(\delta+2\be-\tau)\phi_2-(\Delta+2\mu)\phi_1+\nu\phi_0=\frac{j_2}{2},\nn&&
(\delta-2\tau)\phi_1-(\Delta-2\ga+\mu)\phi_0+\sigma\phi_2=\frac{j_3}{2},\nn&&
(D+2\epsilon-\rho)\phi_2-(\bar{\delta}+2\pi)\phi_1+\lambda\phi_0=\frac{j_4}{2},
\eea
where $D\equiv{}l^\mu\prt_\mu,~\Delta\equiv{}n^\mu\prt_\mu,~\delta\equiv{}m^\mu\prt_\mu,
~\bar{\delta}\equiv{}\bar{m}^\mu\prt_\mu$ are the directional derivatives along the null directions
$l^\mu,~n^\mu,~m^\mu,~\bar{m}^\mu$, respectively,
and the notation of the spin coefficients $\ga_{abc}$ are the same as in Ref. \cite{NP1962-68}.
For example, $\pi=\ga_{421},~\si=\ga_{133},~\rho=\ga_{134}$,\,...\,.
Note that some spin coefficients possess well-defined geometric interpretations.
Take $\sigma,~\rho$ as examples: $\sigma$ represents complex shear of $l^\mu$,
$\rho$ represents the complex expansion of a congruence
(including divergence/convergence and twist) associated with $l^\mu$,
similarly for the other coefficients \cite{Sachs1960s}\cite{NP1962-68}.

\subsubsection{The differential form approach}\label{Diff-form}
To understand the motivation for defining the three complex Newman-Penrose (NP) scalars,
it is helpful to work within the framework of differential forms.
This approach not only streamlines the derivations
but also ensures compactness of expressions and coordinate invariance in its formulation.
In particular, when the relevant theories can be cast in differential form,
it becomes particularly straightforward to derive the NP formulation of Maxwell's equations.

For clarity, let us begin with a simple example: an electromagnetic wave propagating along the $z$-direction
in flat spacetime with Cartesian coordinates.
This setting helps illustrate the basic physical meaning of the NP scalars before turning to the more general case.
In this setup, the family of equal-phase surfaces is given by $u_z\equiv\frac{t-z}{\sqrt{2}}=\mathrm{const.}$,
which defines the retarded time coordinate.
Similarly, the advanced null coordinate is  $v_z\equiv\frac{t+z}{\sqrt{2}}$.
In flat spacetime with Cartesian coordinates $\{t,x,y,z\}$,
the remaining two null coordinates can be defined as $X_\pm\equiv\frac{x\pm iy}{\sqrt{2}}$.
In terms of these null coordinates, the line element becomes $ds^2=2(dX_+dX_--du_zdv_z)$,
and the flat space metric in this coordinate reads
\bea
(\eta_{\mn})=\left(
               \begin{array}{cccc}
                 0 & -1 & 0 & 0 \\
                 -1 & 0 & 0 & 0 \\
                 0 & 0 & 0 & 1 \\
                 0 & 0 & 1 & 0 \\
               \end{array}
             \right).
\eea
Then the set of null tetrad can be defined as $\{E^a_{~\nu},a=1,...,4\}=\{du_z,dv_z,dX_+,dX_-\}$,
and the coordinate 1-forms can be expressed in these null 1-forms as
\bea&&
dt=\frac{du_z+dv_z}{\sqrt{2}},\quad dz=\frac{dv_z-du_z}{\sqrt{2}},\nn&&
dx=\frac{dX_++dX_-}{\sqrt{2}},\quad dy=\frac{i(dX_--dX_+)}{\sqrt{2}},
\eea
With these coordinate 1-forms, we can express the Faraday 2-form as
\bea&&\label{Faraday-2f}
\hspace{-3.6mm}
F=\hf F_{\mn}dx^\mu\wedge dx^\nu=E^i\,dx^i\wedge dx^0+\frac{B^k}{2}\epsilon_{ijk}\,dx^i\wedge dx^j\nn&&
\hspace{-3mm}
=[E^z{\bf{n}\wedge\bf{l}}+i\,B^z{\bf{m}\wedge\bf{\bar{m}}}]
+(E_+{\bf\bar{m}}+E_-{\bf{m}})\wedge({\bf{n}+\bf{l}})\nn&&
\hspace{-3mm}
+i(B_+{\bf\bar{m}}-B_-{\bf{m}})\wedge({\bf{n}-\bf{l}}),
\eea
where for notational simplicity, we have defined $E_\pm\equiv\hf(E^x\pm i\,E^y)$ and $B_\pm\equiv\hf(B^x\pm i\,B^y)$,
and have used the conventional notation $\{{\bf{l}},{\bf{n}},{\bf{m}},{\bf{\bar{m}}}\}=\{du_z,dv_z,dX_+,dX_-\}$.
The above formula is instructive:
the last equation indicates that we can classify $\vec{E},~\vec{B}$ into three groups,
the $z$-component part $E^z, B^z$ parallel to the wave propagation direction
and the transversal parts $E_\pm,~B_\pm$.
Remember that the dual Faraday tensor of $(F_{i0},~F_{ij})=(E^i,\frac{\epsilon_{ijk}}{2}B^k)$
is $({^*F}_{i0},~{^*F}_{ij})=(-B^i,\frac{\epsilon_{ijk}}{2}E^k)$,
we can immediately obtain the dual 2-form
\bea&&\label{dualFaraday-2f}
\hspace{-3.6mm}
{^*F}=\hf{^*F}_{\mn}dx^\mu\wedge dx^\nu
=\frac{E^k}{2}\epsilon_{ijk}\,dx^i\wedge dx^j-B^i\,dx^i\wedge dx^0\nn&&
\hspace{-3mm}
=[i\,E^z{\bf{m}\wedge\bf{\bar{m}}}-B^z{\bf{n}\wedge\bf{l}}]
-(B_+{\bf\bar{m}}+B_-{\bf{m}})\wedge({\bf{n}+\bf{l}})\nn&&
\hspace{-3mm}
+i(E_+{\bf\bar{m}}-E_-{\bf{m}})\wedge({\bf{n}-\bf{l}}).
\eea
Adding Eq. (\ref{Faraday-2f}) and Eq. (\ref{dualFaraday-2f}) together,
we can easily form a complex 2-form
\bea&&\label{Comp-Faraday-2f}
\hspace{-3.6mm}
F_d=\hf(F-i\,{^*F})=(E^z+i\,B^z)({\bf{n}\wedge\bf{l}+\bf{m}\wedge\bf{\bar{m}}})
\nn&&\hspace{-3mm}
+(E_++iB_+){\bf{\bar{m}}\wedge\bf{n}}+(E_-+iB_-){\bf{m}\wedge\bf{l}}.
\eea
Interestingly, by a tedious and trivial calculation,
we can confirm that the three complex scalars defined in (\ref{NPS-Faraday}) are simply
\bea&&\label{PlaneWz}
\phi_0=-(E_++iB_+),\quad \phi_2=(E_-+iB_-),\nn&&
\phi_1=-(E^z+iB^z),
\eea
which are precisely the coefficients of the corresponding basis 2-forms 
({\it e.g. ${\bf{m}\wedge\bf{l}}$}) in the decomposition (\ref{Comp-Faraday-2f}).
Note that we do not distinguish upper and lower spatial indices for pure spatial components in Cartesian coordinates,
such as $E^z=E_z,~B_x=B^x$.
With the simple case shown above, we can see the underlying advantages in defining the three complex NP scalars:
\bit
\item $\phi_1$ represents the longitudinal mode parallel to the wave propagation direction 
[say, $\phi_1\propto\hat{e}_z$ in (\ref{PlaneWz})], and thus can be regarded as the Coulomb mode of EM fields.
\item Both $\phi_0$ and $\phi_2$ describe the transversal parts of EM fields,
although they represent opposite polarization configurations.
In the Jones vector convention, $E_+$ represents right-handed elliptic (or circular) polarization,
while $E_-$ represents the left-handed one.
This can be understood by noting that the $y$-component of $\vec{E}$ field is phase advanced (or retarded) by
$\frac{\pi}{2}$ relative to the $x$-component.
When viewing along the propagation direction $\hat{e}_z$,
this phase difference causes the $\vec{E}_{\pm}$ (and likewise $\vec{B}_{\pm}$) to rotate clockwise or counterclockwise,
corresponding to left- or right-handed polarization, respectively.
\item Parity is implemented by the swaps
$l\leftrightarrow n$ and $m\leftrightarrow-\bar{m}$.
Under this transformation one finds $\phi_0\leftrightarrow+\phi_2$ and $\phi_1\leftrightarrow-\phi_1$,
which interchanges the outgoing and incoming radiation modes, and also flips the chirality through $i\leftrightarrow-i$.
One can then readily verify that this transformation exchanges the equations in Eqs. (\ref{LINPMax-1form}) in pairs,
namely, equations in between $(j_1,~j_2)$ and $(j_3,~j_4)$.
We emphasize that the parity operation is subtle, since one must specify which objects are being transformed.
One may regard the transformation as acting either on the NP scalars $\phi_a$ ($a=0,1,2$), 
\ie, the component of $F_{\mn}$ in a fixed tetrad;
or on the tetrad $\{{\bf{l}},{\bf{n}},{\bf{m}},{\bf{\bar{m}}}\}$ itself.
If parity transformation is applied simultaneously to both objects,
the geometric 2-form $F=\hf{F_{\mn}}dx^\mu\wedge dx^\nu$ remains invariant,
and the overall NP equations are unaffected.
\eit
In short conclusion, the compact form of Faraday 2-form can be reexpressed in terms of the three complex NP scalars,
\bea&&\label{F-2formNull}
\hspace{-3.6mm}F_d=
\phi_1({\bf{l}\wedge\bf{n}-\bf{m}\wedge\bf{\bar{m}}})-\phi_2{\bf{l}\wedge\bf{m}}
+\phi_0{\bf{n}\wedge\bf{\bar{m}}}.
\eea

In the following, we outline the basic logic for deriving the Maxwell equations in NP form.
The LI Maxwell action, written in differential-form language, is
\bea
I_A=\int(-\hf F\wedge{^*F}),
\eea
from which the equation of motion follows:
\bea\label{dFormMaxw}
d(^*F)=-(^*J).
\eea
The homogeneous Maxwell equation, $dF=0$, arises naturally from the definition $F=dA$. 
With some straightforward algebraic manipulations, 
the Maxwell equations can then be cast into NP form.
The key idea is to exploit {\it the coordinate independence of differential forms},
as reflected in Eq. (\ref{dFormMaxw}) and $dF=0$.

We now work directly with the null 1-forms $\bf{E}^a=E^a_{~\mu}dx^\mu$ ($a=1,2,3,4$), 
namely $\{\bf{l},\bf{n},\bf{m},\bf{\bar{m}}\}$,
to derive the equation of motion.
The Faraday 2-form is then written as
$F=\hf F_{ab}\mathbf{E}^a\wedge\mathbf{E}^b$, 
from which the homogeneous Maxwell equation follows:
\bea&&\label{HMENF}
\hspace{-5mm}0=dF=\hf dF_{ab}\wedge\mathbf{E}^a\wedge\mathbf{E}^b+F_{ab}\,d\mathbf{E}^a\wedge\mathbf{E}^b\nn&&
\hspace{-2mm}=\hf[\prt_{c}F_{ab}+2F_{b}^{~d}\ga_{cda}]\mathbf{E}^a\wedge\mathbf{E}^b\wedge\mathbf{E}^c
\Leftrightarrow\nn&&
\nabla_{c}F_{ab}+2F_{b}^{~d}\ga_{cda}=0,
\eea
where we have used the orthogonality relation $E_b^{~\mu}E^a_{~\mu}=\delta^a_b$ to reverse the null 1-form ${\bf{E}}^a$ 
in order to obtain the coordinate 1-form $dx^\mu=E_a^{~\mu}{\bf{E}^a}$. 
In the intermediate steps, we have used $dF_{ab}=\prt_\nu F_{ab}\,dx^\nu=\prt_\nu F_{ab}\,E_c^{~\nu}{\bf{E}^c}=\prt_c F_{ab}\,{\bf{E}}^c$
and $d\mathbf{E}^a=E_c^{~\nu}\nabla_\nu E^a_{~\mu}E_b^{~\mu}{\bf{E}}^c\wedge{\bf{E}}^b=\eta^{ad}\ga_{bdc}{\bf{E}}^c\wedge{\bf{E}}^b$.
Note that $\nabla_{c}F_{ab}=\prt_{c}F_{ab}$, since the frame metric $\eta_{ab}$ is a constant matrix, 
as shown in Eq. (\ref{FrameMetric}).

Similarly, from the inhomogeneous Maxwell equation $d(^*F)=-(^*J)$, we can get
\bea\label{IHMENF}
\nabla_{c}(^*F)_{ab}+2(^*F)_{b}^{~d}\ga_{cda}=-(^*J)_{cab}.
\eea
Substituting Eqs. (\ref{HMENF}) and (\ref{IHMENF}) into the exterior derivative
of the complex self-dual 2-form (\ref{Comp-Faraday-2f}), 
we obtain
\bea\label{SDDEFT}&&
\hspace{-5mm}
dF_d=\hf\left\{\nabla_{c}[F_{ab}-i(^*F)_{ab}]+2[F_{b}^{~d}-i(^*F_{b}^{~d})]\ga_{cda}\right\}
\nn&&
\cdot\mathbf{E}^a\wedge\mathbf{E}^b\wedge\mathbf{E}^c=\frac{+i}{2}(^*J)_{cab}\mathbf{E}^a\wedge\mathbf{E}^b\wedge\mathbf{E}^c.
\eea
We also recall that the spin coefficient $\ga_{cda}\equiv e_c^{~\mu}e_{d\mu;\nu}e_a^{~\nu}$ is related to the spin connection by
$\ga_{abc}=\omega_{\mu\,ab}e_c^{~\mu}=-\ga_{bac}$.

Amazingly, Eq. (\ref{SDDEFT}) {\it turns out to be the compact form} of the four Maxwell equations
in the NP formalism.
Moreover, it encodes the definition of the three independent nonvanishing complex NP scalars
\bea&&\label{phi_Faraday}
{F_d}_{21}=\hf[F_{21}-i(^*F)_{21}]=\hf[F_{34}-i(^*F)_{34}]=\phi_1,\nn&&
{F_d}_{31}=\hf[F_{31}-i(^*F)_{31}]=F_{31}=\phi_0,\nn&&
{F_d}_{24}=\hf[F_{24}-i(^*F)_{24}]=F_{24}=\phi_2,
\eea
as already shown in Eqs. (\ref{NPS-Faraday}).
The remaining components, such as the $[23]$
and $[41]$ components, are zero, namely, 
\bea&&
\hspace{-5mm}\hf[F_{23}-i(^*F)_{23}]
=\hf[F_{23}-i\,\epsilon_{2341}\eta^{43}\eta^{12}F_{32}]=0,\nn&&
\hspace{-5mm}\hf[F_{14}-i(^*F)_{14}]=0.
\eea
Here we have used $\epsilon_{1234}=i$ and $\eta^{21}\eta^{43}=-1$.
This result is expected, since the self-dual complex Faraday tensor fully encodes the information of
the conventional real Faraday tensor and therefore possesses only three independent nonzero degrees of freedom, 
as indicated in Eq. (\ref{phi_Faraday}).

To get NP form equations explicitly, we may take $\mathbf{E}^1\wedge\mathbf{E}^2\wedge\mathbf{E}^3$
(refereed as $[123]$) as an example.
The right hand side of (\ref{SDDEFT}) is
$\frac{i}{2}(^*J)_{312}=\frac{i}{2}\epsilon_{_{3124}}j^4=-\frac{j_3}{2}$,
and the left hand side reads
\bea&&\label{3diff-Forms}
\hspace{-5mm}2[(\nabla_3{F_d}_{12}+\nabla_2{F_d}_{31})+\eta^{ba}({F_d}_{2a}\ga_{3b1}
+{F_d}_{3a}\ga_{1b2}\nn&&\hspace{-5mm}
+{F_d}_{1a}\ga_{2b3})]
-2\eta^{ba}({F_d}_{1a}\ga_{3b2}+{F_d}_{2a}\ga_{1b3}+{F_d}_{3a}\ga_{2b1})
\nn&&\hspace{-5mm}\Rightarrow
[(\Delta+\mu-2\ga)\phi_0-(\delta-2\tau)\phi_1-\si\phi_2]=\frac{-j_3}{2},
\eea
With a simple rearrangement, we can readily recognize it as the third equation of (\ref{LINPMax-1form}).
Similarly, the other three remaining 3-form equations of $[124]$, $[234]$, $[134]$ in Eqs.(\ref{SDDEFT}) 
are the corresponding equations of $j^4,~j^2,~j^1$ in Eqs. (\ref{LINPMax-1form}).

\section{The extensions of Maxwell equations in Null formalism}\label{LVMaxwNull}
Following the same procedure used to obtain the LI Maxwell equations (\ref{LINPMax-1form}),
we can also derive the corresponding null formalism for the extended Maxwell equations,
provided that the action admits a formulation in terms of differential forms.
The LI Maxwell theory offers an ideal example, since its action can be written as
\bea&&
I_{0}=\int \Big[-\hf F\wedge{^*F}+{^*j}\wedge A \Big]\nn&&
~~=\int\sqrt{-g}d^4x \Big[-\frac{1}{4}F_{\mn}F^{\mn}+j^\mu A_\mu \Big],
\eea
where $A=A_\nu dx^\nu$ and $F=\hf F_{\mn}dx^\mu\wedge dx^\nu$, and ${^*F}=\frac{1}{4}\epsilon_{\mn}^{~~\,\al\be}F_{\al\be}dx^\mu\wedge dx^\nu$
and ${^*j}=\frac{1}{3!}\epsilon_{\alpha\beta\gamma}^{~~~~\mu} j_\mu dx^\alpha\wedge dx^\beta\wedge dx^\gamma$ are the Hodge dual of
2-form $F$ and 1-form $j$, respectively.

Interestingly, if we replace $F\wedge{^*F}$ by $F\wedge{F}$, we immediately find that
it reduces to a boundary term $\int F\wedge{F}=\int dA\wedge{F}=\int[d(A\wedge{F})+A\wedge{dF}]=\int d(A\wedge{F})$,
and thus does not contribute to the equation of motion.
However, once a scalar function is inserted, things can be quite different.
In that case, we either obtain Lorentz-invariant axion electrodynamics or 
the CPT-odd Maxwell–Chern–Simons (MCS) theory \cite{SMEa}
(also known as the CFJ theory \cite{CFJ1990} in the literature).
This simple example illustrates how differential forms naturally facilitate the construction and analysis of both Lorentz-invariant and Lorentz-violating extensions. In the following subsections, we systematically apply this differential-form framework to several representative Lorentz-violating extensions, beginning with the CPT-odd Chern–Simons term and then proceeding to CPT-even and higher-dimensional operators.

\subsection{The CPT-odd MCS/CFJ theory}\label{d3LVMaxw}
Let us first consider the MCS theory, which modifies the standard Maxwell electrodynamics 
by introducing a CPT-odd coupling between a scalar function/field and the Chern-Simons term $F\wedge{F}$ in the action. 
In four-dimensional spacetime, the Chern-Simons term $F\wedge{F}$ is a topological invariant and therefore, 
by itself, does not contribute to the equation of motion, as mentioned above. 
A nontrivial contribution arises once it is coupled to a scalar function $\theta$, 
in which case one arrives at
\bea&&\label{CFJ-form}
\delta I_O\supset\int-\hf\theta(x) F\wedge{F}=\int-\hf\theta(x) dA\wedge{F}\nn&&
=-\int\hf\Big\{d\Big[\theta(x)A\wedge{F}\Big]-d\theta\wedge A\wedge{F}+\theta(x)A\wedge{dF}\Big\}\nn&&
=\hf\int d\theta\wedge A\wedge{F}-\hf\int d\Big[\theta(x)A\wedge{F}\Big],
\eea
where the third term in the second line vanishes due to $dF=0$, and the last term in the last line is a boundary term
and can be ignored.
Identifying $d\theta\rightarrow-2(k_{AF})$, we can arrive at the CFJ theory \cite{CFJ1990},
or the CPT-odd MCS theory \cite{SMEa},
\bea\label{CFJAct}
I_\mathrm{CFJ}=I_{0}+\int \big[A\wedge F\wedge (k_{AF})\big],
\eea
where $k_{AF}\equiv(k_{AF})_\mu dx^\mu$.
The equation of motion in differential form is
\bea\label{CPT-odd-dF}
d{^*F}-2(k_{AF})\wedge F=-{^*j}.
\eea
Note that the superficially non-gauge-invariant action in (\ref{CFJAct}) changes 
under the transformation $A\rightarrow A+d\Lambda$
only by a surface term of the form $\Lambda(x)\,F\wedge (k_{AF})$, 
where $\Lambda(x)$ is an arbitrary scalar function.
This follows from the identities $dF=0$ and $d(k_{AF})=0$.
The theory therefore remains gauge invariant up to boundary contributions,
as can be verified in Eq. (\ref{CPT-odd-dF}).
In the axion scenario, 
the Lagrangian may be completed by including the pseudo-scalar dynamics
\bea\label{axion-UVcomp}
\mathcal{L}_\theta=-\hf\nabla\theta\cdot\nabla\theta-V[\theta]
\eea
so that the gradient of the scalar field $\theta$ can acquire a preferred value determined by the prescribed potential $V[\theta]$.
In differential-form language, $\hf\int d\theta\wedge({^*d\theta})=-\hf\int\sqrt{-g}d^4x\nabla\theta\cdot\nabla\theta$,
which closely resembles the structure appearing in dynamical Chern-Simons modified gravitational theory \cite{CSG2009}.
It is worth noting, however, that a natural Chern-Simons theory fundamentally lives in 2+1 dimensions (or odd dimensions higher than three, such as the five-dimensional Einstein-Maxwell-Chern-Simons theory \cite{EMCSZL}).
A genuine electromagnetic Chern-Simons theory formulated in terms of differential forms has already been realized
in planar electrodynamics \cite{Marco2023}.

We also want to point out that a genuine Chern-Simons theory or the action (\ref{CFJ-form}) does not need the metric tensor,
in contrast to the case of Maxwell theory or dynamical Chern-Simons, which requires the Hodge duality operation, ``$*$",
and thus cannot avoid the need for a metric tensor.
This distinction is important because in pre-metric theory of electrodynamics \cite{SMFLE1999}\cite{SMLE2001}, the metric tensor is ``derived".
In other words, the metric tensor is not assumed at the outset,
but rather as a secondary quantity ``derived" from the constitutive tensor $\chi_{\mn}^{~~\alpha\beta}$ as a link
between excitation tensor $H=(\mathcal{D},\,\mathcal{H})$ and field strength tensor $F=(E,\,B)$, \ie, 
$H_{\mn}=\chi_{\mn}^{~~\alpha\beta}F_{\alpha\beta}$ \cite{SMFLE1999}\cite{SMLE2001}.
Moreover, the pre-metric electrodynamics is also well-suited to be described in differential forms,
and it naturally incorporates axion and skewon electrodynamics \cite{Itin0709}.

A discussion of the MCS theory in the null formalism can be found in Ref. \cite{ZX2025}.
For completeness, we summarize the relevant results here. 
We define $k^a\equiv(k_{AF})_\mu$ not to be confused with the wave vector) for notational simplicity, 
and treat it as a fixed background covector. 
The inhomogeneous Maxwell equations in the null basis then reduce to
\bea\label{CPT-O-Maxw}&&
\eta^{ac}F_{ab|c}+\epsilon_{abcd}\,k^aF^{cd}=-(j_e)_b.
\eea
Furthermore, we adopt the test particle assumption and 
assume that the background geometry is the Minkowskian or Schwarzschild geometry for simplicity,
then the NP equations are given by
\begin{widetext}
\bseq\label{CPTO-NP-Maxwell}
\bea&&
(\delta+2\be-\tau)\phi_2-(\Delta+2\mu)\phi_1+\nu\phi_0=\frac{j_2}{2}+2k^1\mathrm{Im}[\phi_1]+ik^3\bar{\phi}_2
-ik^4\phi_2,\label{CPTO-NP-2}\\&&
(\delta-2\tau)\phi_1-(\Delta-2\ga+\mu)\phi_0+\sigma\phi_2=\frac{j_3}{2}-ik^1\phi_0-2ik^4\mathrm{Re}[\phi_1]
-ik^2\bar{\phi}_2,\label{CPTO-NP-3}\\&&
(D+2\epsilon-\rho)\phi_2-(\bar{\delta}+2\pi)\phi_1+\lambda\phi_0=\frac{j_4}{2}+ik^1\bar{\phi}_0
+2ik^3\mathrm{Re}[\phi_1]+ik^2\phi_2.\label{CPTO-NP-4}\\&&
(D-2\rho)\phi_1-(\bar{\delta}-2\al+\pi)\phi_0+\kappa\phi_2=\frac{j_1}{2}+ik^3\phi_0-ik^4\bar{\phi}_0
-2k^2\mathrm{Im}[\phi_1],\label{CPTO-NP-1}
\eea
\eseq
\end{widetext}
In flat spacetime, we have
\bea&&
\hspace{-5mm}{l}^\mu=\delta^\mu_1\Rightarrow D=\prt_r,\nn&&
\hspace{-5mm}n^\mu=\delta^\mu_0+U\delta^\mu_1+X^A\delta^\mu_A \Rightarrow \Delta=[\prt_u-\frac{\prt_r}{2}],\nn&&
\hspace{-5mm}m^\mu=\frac{1}{\sqrt{2}}(\delta_2^\mu+i\csc\theta\delta^\mu_3)\Rightarrow\delta=\frac{1}{\sqrt{2}\,r}[\prt_\theta+\frac{i\,\prt_\phi}{\sin\theta}],\nn
\eea
where the parameters are given by $X^A=\omega=0$, $U=-\hf$, and $\xi^A=\frac{1}{\sqrt{2}r} (\delta_2^A + i \csc \theta \delta^A_3 )$.

Inspection of the above equations reveals a structural similarity to the LI Maxwell equations.
The coupled system naturally separated into two pairs:
the last pair, Eqs. (\ref{CPTO-NP-4}) and (\ref{CPTO-NP-1}), contain no time derivatives and therefore act as constraint equations on a given hypersurface $u=\mathrm{const.}$.
In contrast, the first pair, Eqs. (\ref{CPTO-NP-2}) and (\ref{CPTO-NP-3}) involve time derivatives $\prt_u\phi_0$ and $\prt_u\phi_1$,
and thus serve as dynamical equations governing the evolution of photon fields away from a given null hypersurface.

It is important to note that there is no time derivative of $\phi_2$, \ie, no $\prt_u\phi_2$ term present in the above equations.
Consequently, $\phi_2(u,\theta,\phi)$ remains the dynamical radiation degree of freedom.
This is consistent with the two radiation degrees of freedom of conventional LI Maxwell theory (when counting only real degrees of freedom),
since our modification still describes a massless spin-1 photon field via the gauge potential $A_\mu$, preserving gauge invariance.

A notable feature of these coupled equations is that the CPT-odd coefficients $k^a$ couple the three NP scalars together,
as evident from the right-hand side of Eqs. (\ref{CPTO-NP-Maxwell}).
This coupling renders the system considerably more intricate.
The right-hand side may be interpreted as an effective current, analogous to the external source $\frac{j_a}{2}$.
Consequently, even in vacuum where the external current vanishes ($j_a=0$), 
the induced ``source" generated by the background electromagnetic fields remains nonzero, 
thereby producing a nonzero magnetic field even in the absence of external source \cite{LVEM}.
In this case, the system closely resembles the axion electrodynamics, since $k_a=\nabla_a\theta$ can be identified as the gradient of the axion field.

\subsection{A CPT-even anisotropic extension}\label{d4LVMaxw}

In this subsection, we turn to consider the CPT-even anisotropic extension of Maxwell theory, 
which incorporates a dimension-four, CPT-even operator in the action. 
The relevant term takes the form $(k_F)^{\mn\al\be}F_{\mn}F_{\al\be}$ \cite{SMEa},
where $(k_F)^{\mn\al\be}$ may be viewed as a subset of the constitutive tensor $\chi^{\mn\alpha\beta}$,
with the totally antisymmetric axion part proportional to
$\epsilon^{\mn\al\be}$ and the skewon part $\chi^{\mn\al\be}=-\chi^{\al\be\mn}$ \cite{Itin0709} excluded.
The tensor $(k_F)^{\mn\al\be}$ contains 20 independent parameters and is therefore rather cumbersome to handle. 
To our knowledge, this theory was first introduced in the minimal SME \cite{SMEa},
and was further discussed in Ref. \cite{SMEbmini}.
Extensions to arbitrary dimensions and additional theoretical issues can be found in Ref. \cite{SMEb, SMEgauge, nonminiEM}.
In what follows, we restrict our attention to the dimension-four $k_F$ term and denote it by $\kappa$ for notational simplicity.

From the pre-metric viewpoint \cite{SMFLE1999, SMLE2001},
the coefficients $\kappa^{\mn\al\be}$ correspond to the principal part of the constitutive tensor $\kappa^{A}_{~B}$,
where $A,~B$ label antisymmetric index pairs $[\mn]$ and $[\rho\sigma]$.
One may thus regard $\kappa^{A}_{~B}$ as a linear map on 2-forms, $\kappa:\Omega^2(\mathcal{M})\mapsto\Omega^2(\mathcal{M})$.
so that pre-metric linear electrodynamics admits a compact coordinate-free formulation in terms of differential forms. 
While the pre-metric approach to electrodynamics and gravity \cite{SMFLE1999, SMLE2001} is itself a rich subject, 
here we focus only on the $\kappa^{\mn\al\be}$ term,
which shares the symmetries of the Riemman tensor, and can be decomposed into Ricci-like and Weyl-like pieces,
denoted by $\tilde{R}_{\mn}$ and $\tilde{W}_{\mn\rho\sigma}$,
respectively (the tilde indicates that these are LV coefficients rather than true curvature tensors). 
They can then be projected into null scalars, in analogy with 
$\Lambda$ and $\Phi_{ab}$ ($a,b=0,1,2$), and $\Psi_a$ ($a=0,...,4$) in the NP formalism \cite{NP1962-68, ZX2025}.

The NP-form Maxwell equations for a special choice of the $\kappa$ term,
where only $\Psi_2=\tilde{W}_{\mn\rho\sigma}l^\mu m^nu\bar{m}^\rho n^\sigma\neq0$ 
were explicitly presented in Eq. (49) of Ref. \cite{ZX2025}.
In principle, the NP equations can be derived for the full set of the coefficients $\kappa_{\mn\alpha\beta}$
by viewing $\kappa$ as a linear map on 2-forms. Written in component form, this map reads
\bea\label{LMap2b2}
F_{\mn}\rightarrow {^{\kappa}F}_{\mn}\equiv\hf\kappa_{\mn\alpha\beta}\,
g^{\alpha\rho}g^{\beta\sigma}F_{\rho\sigma}.
\eea
The corresponding correction to the action is
\bea
\hspace{-5mm}-\frac{1}{4}\int\sqrt{-g}\,d^4x\,\kappa^{\mn\alpha\beta}F_{\mn}F_{\alpha\beta}
=-\int(^*F)\wedge(^{\kappa}F),
\eea
and the inhomogeneous equation of motion in differential form reads
\bea\label{kFDiff}
d[^*F+2{^*(^{\kappa}F)}]+^*j=0,
\eea
where we assume that the variation acts only on the dynamical photon field, so that
$\delta \kappa^{\mn\alpha\beta}=\delta g_{\mn}=0$.
Consequently, $[^*,\delta]=0$,
since the Hodge dual depends only on the metric. 
Under these assumptions, we have also used the identity
$\int(^{\kappa}F)\wedge{^*\delta F}=\int{^*F}\wedge(^{\kappa}\delta F)$
in deriving Eq. (\ref{kFDiff}).
Projecting Eq. (\ref{kFDiff}) into the NP formalism involves only straightforward algebraic manipulations. 
However, the calculation is tedious and the resulting expressions are lengthy, 
so we do not present them explicitly here.

However, if we restrict ourselves to a birefringence-free subset of the CPT-even coefficients, 
as parameterized in Ref. \cite{Marco2012}, the relevant correction takes the form
\bea&&\label{PO-CPTE-photon}
-\frac{1}{4}\kappa^{\mn\al\be}F_{\mn}F_{\al\be}\supset
-\frac{1}{4}[({c}_F)^{\lambda\nu}F^\mu_{~\lambda}-({c}_F)^{\kappa\mu}F_{~\kappa}^{\nu}]F_{\mn}\nn&&
\supset-\frac{1}{2}\xi^{(\rho}\zeta^{\sigma)}F_{\rho\mu}F_\sigma^{~\mu},
\eea
where we have used the parametrization \cite{Marco2012, FHRL2016}
\bea
({c}_F)^{\rho\sigma}=\xi^{(\rho}\zeta^{\sigma)}\equiv\hf[\xi^{\rho}\zeta^{\sigma}+\xi^{\sigma}\zeta^{\rho}].
\eea
With this restriction, the explicit NP equations can be explicitly presented in Eqs. (\ref{CPT-E-cNP}).
The last term in Eq. (\ref{PO-CPTE-photon}) can be written in differential form as
\bea
-\hf\int[i_\xi F\wedge{^*(i_\zeta F)}+i_\zeta F\wedge{^*(i_\xi F)}]\subset\delta I_E,
\eea
where $\xi\equiv\xi^\rho\prt_\rho$ and $\zeta\equiv\zeta^\rho\prt_\rho$ are background vector fields on the tangent space $T\mathcal{M}$
with $\mathcal{M}$ the spacetime manifold,
and $i_\xi F=\xi^\rho F_{\rho\sigma}dx^\sigma$ denotes the interior contraction of $\xi$ with the 2-form $F$ (similarly for $i_\zeta F$).
Varying only the dynamical photon field and keeping the background vectors and metric fixed 
($\delta\zeta=\delta\xi=\delta g=0$),
the equation of motion becomes
\bea\label{PO-CPTE-Maxwell}
d[(^*F)+\hf(i_{\zeta}(^*i_\xi F)+i_{\xi}(^*i_\zeta F)]+(^*j)=0.
\eea
From Eq. (\ref{PO-CPTE-Maxwell}), a short algebraic manipulation yields the component form
\bea\label{Latin-CPTE-Maxwell}
F^{cd}_{~~|d}+\hf[F^{c}_{~m}\xi^{(d}\zeta^{m)}-F^{d}_{~m}\xi^{(c}\zeta^{m)}]_{|d}=j^c,
\eea
which coincides with the spacetime-index expression derived directly from the Maxwell Lagrangian
with correction (\ref{PO-CPTE-photon}), namely,
\bea\label{Greek-CPTE-Maxwell}
\nabla_\nu[F^{\mn}+\hf \xi^{(\mu}\zeta^{\rho)}F^\nu_{~\rho}-\hf \xi^{(\nu}\zeta^{\rho)}F^\mu_{~\rho}]=j^\mu.
\eea
In fact, Eqs. (\ref{Greek-CPTE-Maxwell}) and (\ref{Latin-CPTE-Maxwell}) represent the same tensor equation 
expressed in spacetime manifold form and tangent space form, respectively.
This equivalence follows from a simple relation between the spacetime covariant derivative 
and the intrinsic (frame) derivative of a given tensor, namely,
\bea\label{intrin-covarD}
\nabla_\ga\,T^{\al_1...\al_n}_{~~\be_1...\be_m}
=e_{a_1}^{~\al_1}...e_{~a_n}^{\al_n}e^{b_1}_{~\be_1}...e^{b_m}_{~\be_m}
e^c_{~\ga}T^{a_1...a_n}_{~~b_1...b_m|c}.
\eea
However, the situation is more subtle for non-dynamical background fields such as $\xi^\rho$ or $\zeta^\rho$.
In general, the covariant derivative of a background vector field need not vanish, even in Cartesian coordinate, 
unless the spacetime is flat or one works in a special class of freely falling frames 
({\it e.g.}, the Sun's free-fall frame with respect to the Galaxy, 
or approximately inertial motion in weak gravity).
Generically, even in arbitrary Fermi normal coordinates (FNC), 
the relation
\bea\label{covDbkg}
\nabla_\mu\xi^\nu=\prt_\mu\xi^\nu
+\Gamma^\nu_{\mu\rho}\xi^\rho
\xlongequal[\text{FNC}]{\Gamma^\alpha_{\mn}=0}
\prt_\mu\xi^\nu=0
\eea
is not necessarily valid, though the last equality holds true in the Cartesian coordinate without LV.
The condition that a generic LV tensor coefficient has a vanishing covariant derivative cannot, 
in general, be satisfied in an arbitrary spacetime, as has been discussed extensively in Ref. \cite{SMEg},
except in special trivial cases such as a parallelization manifold or in an asymptotic region
when it is imposed as a boundary condition \cite{SMEg}.
Interestingly, it is precisely the asymptotic region that is relevant for our purposes, 
since it is our main concern about the asymptotic behavior of the LV-corrected Maxwell equations \cite{ZX2025}.
Therefore, it is at least consistent to impose $\nabla_\mu\xi^\nu|_\gamma=0$
along the world curve $\gamma$ of a specific freely falling observer.

With this assumption in place,
Eq. (\ref{intrin-covarD}) implies
$\xi^a_{~|b}=e^a_{~\mu}e_b^{~\nu}\xi^\mu_{;\nu}$
and hence $\xi^a_{~|b}|_\gamma=0$.
This result substantially simplifies our subsequent calculations.
However, 
it is important to keep in mind that
the intrinsic derivatives
\bea\label{intrinbkg}
\xi^a_{~|b}=\xi^a_{~,b}-\eta^{ad}\xi^m\gamma_{mdb}
\eea
do not, in general, vanish in a small neighborhood of a generic spacetime point, nor away from the world line $\gamma$.
Indeed, even the first term, the directional derivative along $e^a_{~\mu}$, 
$\xi^a_{~,b}=e^a_{~\mu}\xi^\mu_{~b}$, 
need not be zero, let alone the full intrinsic derivative $\xi^a_{~|b}$.

Although assuming $(\xi^a_{~|b})|_\gamma=(\xi_{a|b})|_\gamma=0$ within a suitably chosen class of FNCs 
greatly simplifies the analysis,
it is also reasonable to consider an alternative, yet still viable assumption.
Suppose the background vectors $\xi^\mu,\zeta^\mu$ are timelike.
One may then exploit this property to impose
$\xi^{A}=e^A_{~\mu}\xi^\mu=0=\zeta^{A}=e^A_{~\mu}\zeta^\mu$ with $A=3,4$,
since the projections of time-like vectors $\xi^\mu,~\zeta^\mu$ onto intrinsically
spacelike directions $m^\mu,~\bar{m}^\mu$ vanish within a suitably restricted class of frames.

A remaining issue is whether such restricted frames have any nonzero overlap with the specific FNC
constructed along a given world curve $\gamma$.
This overlap is not guaranteed {\it a priori}.
The complex null vectors $m^\mu,~\bar{m}^\mu$ are constructed from spacelike tetrads associated with a null congruence,
and the corresponding null tetrads behave as dynamical fields, transforming non-trivially under local Lorentz transformations, just as $F_{\mn}$.
Consequently,
the conditions $\xi^{A}=\zeta^{A}=0$ can be satisfied only in
specially chosen frames
and are not preserved under general (in particular, dynamical) Lorentz transformations.

This frame dependence encapsulates the essential feature of Lorentz violation 
induced by background tensor fields within the SME framework.
Nevertheless, conditions $\xi^{A}=\zeta^{A}=0$ may be preserved under observer Lorentz transformations,
provided that they can be imposed in a particular frame in certain specific spacetimes.

Now we are in a position to introduce the CPT-even correction
\bea&&
-\frac{i}{2}d[i_{(\zeta}(^*i_{\xi )}F)]
=-\frac{i}{4}\epsilon_{bcd}^{~~~m}[\xi^{(d}\zeta^{n)}F_{nm}]_{|a}[abc]
\eea
into the LI Maxwell equation
$dF_d-\frac{i}{2}^*j=0$, whose components equations in NP-form are given by Eqs. (\ref{LINPMax-1form}).
For simplicity, we denote $[abc]\equiv{\bf{E}}^a\wedge{\bf{E}}^b\wedge{\bf{E}}^c$.
The full component forms of the Maxwell equation
\bea
dF_d-\frac{i}{2}d[(i_{(\zeta}(^*i_{\xi )}F)-\frac{i}{2}{^*j}=0
\eea
are very complicated, and the calculation is quite tedious.
However, by assuming vanishing angular projections $\xi^A=\zeta^A=0$ ($A=3,4$) 
and within specific FNC frames (if allowed),
the calculation of $-\frac{i}{4}\epsilon_{bcd}^{~~~m}F_{nm}[\xi^{(d}\zeta^{n)}]_{|a}$
can be avoided.
The CPT-even Maxwell equations in NP-component form are then given by
\bseq\label{CPT-E-cNP}
\bea&&
(D-2\rho)\phi_1-(\bar{\delta}-2\al+\pi)\phi_0+\kappa\phi_2-\frac{j_1}{2}
=\nn&&
\mathrm{Re}\left(2[D\phi_1+\kappa\phi_2-\pi\phi_0]-[(\bar{\delta}-2\alpha)\phi_0+2\rho\phi_1]\right)\xi^{(1}\zeta^{2)}\nn&&
+\mathrm{Re}[(\delta+2\be)\phi_2-2\mu\phi_1]\xi^2\zeta^2,\label{CPT-E-NP1}\\&&
(\delta+2\be-\tau)\phi_2-(\Delta+2\mu)\phi_1+\nu\phi_0-\frac{j_2}{2}
=\nn&&
\mathrm{Re}\left([(\delta+2\be)\phi_2-2\mu\phi_1]-2[\Delta\phi_1+\tau\phi_2-\nu\phi_0]\right)\xi^{(1}\zeta^{2)}\nn&&
-\mathrm{Re}[(\bar{\delta}-2\alpha)\phi_0+2\rho\phi_1]\xi^1\zeta^1,\label{CPT-E-NP2}\\&&
(\delta-2\tau)\phi_1-(\Delta-2\ga+\mu)\phi_0+\sigma\phi_2
-\frac{j_3}{2}
=\nn&&
\left([(D+2\bar{\epsilon})\bar{\phi}_2-2\bar{\pi}\bar{\phi}_1]-[(\Delta-2\gamma)\phi_0+2\tau\phi_1]\right)\frac{\xi^{(1}\zeta^{2)}}{2}+\nn&&
\left([(\Delta+2\bar{\gamma})\bar{\phi}_2-2\bar{\nu}\bar{\phi}_1]\frac{\xi^2\zeta^2}{2}-[(D-2\epsilon)\phi_0+2\kappa\phi_1]\frac{\xi^1\zeta^1}{2}\right),\nn\label{CPT-E-NP3}\\&&
(D+2\epsilon-\rho)\phi_2-(\bar{\delta}+2\pi)\phi_1+\lambda\phi_0-\frac{j_4}{2}
=\nn&&
\left([(D+2\epsilon)\phi_2-2\pi\phi_1]-[(\Delta-2\bar{\gamma})\bar{\phi}_0+2\bar{\tau}\bar{\phi}_1]\right)
\frac{\xi^{(1}\zeta^{2)}}{2}+\nn&&
\left([(\Delta+2\gamma)\phi_2-2\nu\phi_1]\frac{\xi^2\zeta^2}{2}-[(D-2\bar{\epsilon})\bar{\phi}_0+2\bar{\kappa}\bar{\phi}_1]\frac{\xi^1\zeta^1}{2}\right),\label{CPT-E-NP4}\nn
\eea
\eseq
where the left hand side (first line) of each equation is the LI contribution, see Eqs. (\ref{LINPMax-1form}).
Again, we find that the above equations are invariant under
the simultaneous exchange of null tetrads $l\leftrightarrow n$ and $m\leftrightarrow\bar{m}$,
and the LV corrections in the last two equations involving angular currents, $j_3$ and $j_4$, 
are complex conjugates to each other.
In fact, the two complex equations are not independent,
just as the corresponding LI Maxwell equations.
The asymptotic behavior of the dimension-3 CPT-odd modified photon field was analyzed in Ref. \cite{ZX2025}.
Compared with Eqs. (\ref{CPTO-NP-Maxwell}), 
the LV corrections above involve not only products of spin coefficients with the photon NP scalars $\phi_a$ ($a=0,1,2$)
but also their null directional derivatives. 
Hence, the CPT-even dimension-4 operators may induce qualitatively different modifications to the photon’s asymptotic behavior.
This may be analogous to the fact that the dimension-4 $(k_F)_{\mn\alpha\beta}$ operator
does not preserve photon flux, while the presence of dimension-3 $(k_{AF})_\kappa$ operator itself does \cite{ZX2025}.
It is therefore worthwhile to investigate their distinct effects on both the asymptotic and polarization properties of electromagnetic fields from astrophysical sources.

\subsection{The Myers-Pospelov extension}\label{d5HDE}

The original Myers-Pospelov extension of Maxwell theory is a vector sub-theory
of the dimensional-five kinematic extension
of a general theory with background time-like vector $n^\mu=(1,\vec{0})$ \cite{MPT03}.
This theory (a). is quadratic in the photon field $A_\mu$ and hence represents only a kinematic extension;
(b). contains one additional derivative, making it a dimension-five operator with linear 
corrections to the dispersion relation; and 
(c). is not reducible to lower-dimension operators, 
leading to order $\mathcal{O}(E^3)$ corrections instead of $\mathcal{O}(E\,m^2)$ or $\mathcal{O}(E^2\,m)$.
This CPT-odd theory can be cast in a form analogous to the CFJ theory by identifying
\bea
(k_{AF})^\mu=-\frac{\zeta}{M_\mathrm{Pl}}n^\mu(n\cdot\prt)^2=-\frac{\zeta}{M_\mathrm{Pl}}n^\mu\,n^\rho\,n^\sigma\prt_\rho\prt_\sigma,
\eea
and thus is similar to $(\hat{k}_{AF}^{(5)})^\mu\equiv(\hat{k}_{AF}^{(5)})^{\mu\rho\sigma}\prt_{\rho\sigma}$ coefficients
with $(\hat{k}_{AF}^{(5)})^{\mu\rho\sigma}=-\frac{\zeta}{M_\mathrm{Pl}}n^\mu\,n^\rho\,n^\sigma$ \cite{SMEb, SMEgauge}.

However,  this theory cannot be ultraviolet complete
because the higher-than-quadratic derivative structure introduces ghost modes -- 
a common feature of high derivative theories \cite{HDO13}.
The appearance of superluminal velocities also signals potential causality violations, 
leading to stability and causality issues, which are also central topics in Lorentz-violating theories \cite{stabilty-Ralf05, Reyes10}.
Quantization may further induce negative-norm states and non-unitarity
problem \cite{Reyes08, Reyes10}.
These stability and causality issues can be cured by choosing a spacelike background vector with $n^2>0$
(for signature $+2$) \cite{Reyes10},
which partly motivates the intensive explorations of Horava-Lifschitz theories \cite{Horava09, AZW17},
where a dynamical anisotropic scaling between space and time ensures 
that Lorentz symmetry emerges only in the infrared.
In contrast, for light-like and time-like $n^\mu$, 
the theory typically exhibits runaway modes or suffers from instability and unitarity issues.


In differential forms, the Myers-Pospelov (or extended MCS/CFJ) action can be written compactly as
\bea&&\label{DfMP}
\hspace{-8mm}
\hf\int[i_\xi F\wedge{^*[\mathcal{L}_\xi(i_\xi{^*F})]}-(i_\xi{^*F})\wedge{^*\mathcal{L}_\xi(i_\xi F)}]\subset\delta I_O,
\eea
where to absorb the prefactor $\frac{\zeta}{M_\mathrm{Pl}}$ in the Myers-Pospelov theory, 
we introduce $\xi=\xi^\mu\prt_\mu$ in place of $n=n^\mu\partial_\mu$ to denote the fixed background vector, 
\ie, $\xi^\mu=(\frac{\zeta}{M_\mathrm{Pl}})^{1/3}n^\mu$.
The background vector field $\xi^\mu$ may be interpreted as the dual of a condensate of 
the gradient of an unknown UV scalar field $\Lambda(x)$, $\xi^\mu=g^{\mn}\langle\nabla_\nu\Lambda(x)\rangle$.
This resembles the relation $k_{AF}^\mu=\frac{g^{\mn}}{2}\partial_\nu \theta(x)$,
suggesting a similar UV completion mechanism as for $k_{AF}$
\cite{Bret2020}, see Eq. (\ref{axion-UVcomp}).
By varying the Myers-Pospelov action (\ref{DfMP}) and ignoring the boundary terms, we can immediately
obtain the 3-form equation of motion as below
\bea\label{MP-3form-eom}
\hspace{-3mm}d\left[{^*F}+
[\{i_\xi{^*},\,\mathcal{L}_\xi\,i_\xi{^*}\}-\{{^*}i_\xi,\,{^*}\mathcal{L}_\xi\,i_\xi\}]\frac{F}{2}\right]+{^*j}=0.
\eea
In the FNC, the dimension-5 corrections reduce to
\bea&&
\hspace{-6mm}[\{i_\xi{^*},\,\mathcal{L}_\xi\,i_\xi{^*}\}-\{{^*}i_\xi,\,{^*}\mathcal{L}_\xi\,i_\xi\}]\frac{F}{2}
\xlongequal[]{\text{FNC}}\nn&&
\hspace{-3mm}-\left[ {2} \xi^\rho\xi_{[\mu}(\xi\cdot\nabla)F_{\nu]\rho}+\frac{\xi^2}{2}(\xi\cdot\nabla)F_{\mn}\right]dx^\mu\wedge dx^\nu.
\eea
Thus, it corrects the Lorentz-invariant Maxwell 3-form equation $dF_d-\frac{i}{2}^*j=0$ by the term
\bea\label{MP-3form-corr}
{i}\left[\xi^d\xi_{[a}(\xi\cdot\nabla)F_{b]d}\right]_{|c}[abc],
\eea
since the contribution proportional to $\frac{\xi^2}{2}(\xi\cdot\nabla)F_{ab|c}[abc]$ in FNC vanishes 
due to the Bianchi identity $F_{[ab|c]}=0$.
One can then readily verify that Eq. (\ref{MP-3form-eom}) in FNC reduces to 
the flat spacetime Maxwell equation with the Myers-Pospelov correction, 
\bea
\prt_\mu F^{\mn} + j^\nu + {2}\xi_\mu(\xi\cdot\nabla)^2{^*F}^{\mn}=0.
\eea
Using Eqs. (\ref{MP-3form-eom}) and (\ref{MP-3form-corr}), one can straightforwardly derive 
the corresponding first-order Maxwell equations in the Newman-Penrose formalism.
However, the resulting expressions are lengthy, 
and the nonrenormalizable higher-dimensional operators do not significantly affect the large-distance behavior of the photon field. 
We therefore do not present the explicit equations. 
The inclusion of these higher-dimensional operators mainly serves to show that the procedure for 
translating Maxwell equations from differential-form language to the NP formalism remains valid 
for a broad class of extended Maxwell theories.

As a further example, we may cast the dimension-5 $k^{(5)\alpha\mn\kappa\lambda}$ operator (Table III in \cite{SMEgauge}) 
into the differential forms in flat spacetime, 
\bea\label{k5decomp}
\int [^*(\mathcal{L}_\xi F)]\wedge [^\chi F]=\int \sqrt{-g}d^4x\, \chi^{\mn\kappa\lambda}\xi^\alpha F_{\mn}\,\partial_\alpha F_{\kappa\lambda}, 
\eea
upon identifying $k^{(5)\alpha\mn\kappa\lambda}=\xi^\alpha\chi^{\mn\kappa\lambda}$, 
where $^\chi F$ is defined in analogy with $^\kappa F$ in Eq. (\ref{LMap2b2}) 
as a linear map from 2-forms to 2-forms.
Note $\chi^{\mn\kappa\lambda}=\chi^{[[\mn][\kappa\lambda]]}$ 
is antisymmetric under interchange of the antisymmetric index pairs, $[\mn]$ and $[\kappa\lambda]$.
In this respect, it resembles the skewon tensor \cite{Itin0709}.
In a similar way, we may also put $k^{(6)\alpha\beta\mn\kappa\lambda}$ term \cite{SMEgauge}
into differential form provided 
\bea\label{k6decomp}&&
\int [^*(\mathcal{L}_{(\zeta}\mathcal{L}_{\xi)} F)]\wedge [^\kappa F]=\nn&&
\int \sqrt{-g}d^4x\, \kappa^{\mn\kappa\lambda}\zeta^{(\alpha}\xi^{\beta)} F_{\mn}\,\partial_{\alpha\beta} F_{\kappa\lambda}, 
\eea
where we assume the LV coefficients to be constant observer tensors in flat spacetime,
and identify $k^{(6)\alpha\beta\mn\kappa\lambda}=\zeta^{(\alpha}\xi^{\beta)}\kappa^{\mn\kappa\lambda}$.
Since this contribution corresponds to a dimension-6 operator, 
we next proceed to formulate the extended Maxwell theory to include operators up to mass dimension six, 
in the next subsection.


\subsection{High derivative electrodynamics}\label{d6HDE}

Higher derivative electrodynamics can be traced back to the early 1940s, 
most notably to the pioneering works of Bopp \cite{Bopp} and Podolsky \cite{Podolsky42}.
Among the well studied higher-derivative extensions of Maxwell theory 
are the Bopp–Podolsky (BP) model and Lee–Wick (LW) electrodynamics \cite{LeeWickMaxw}.
The BP theory was originally proposed to remove the infinities 
associated with the self-energy of classical point charges 
while preserving the linearity of the field equations. 
In contrast, the Lee–Wick model was primarily introduced to address ultraviolet divergences 
at the quantum level and improve renormalizability while preserving unitarity, 
albeit at the cost of mild acausality. For concise reviews, see Refs. {\cite{Stability-highD, Marco2023}}.

At the classical level in flat spacetime,
the LW operator $\frac{1}{4}\prt_\mu F_{\alpha\beta}\prt^{\mu}F^{\alpha\beta}$ 
and the BP operator $\hf\prt_\mu F^{\mu\nu}\prt^{\rho}F_{\rho\nu}$ are equivalent up to a total derivative, 
by virtue of the Bianchi identity. 
To express higher-derivative operators in differential forms, 
it is convenient to introduce the operator $^\star\equiv{^*d\,{^*}}$,
which maps a $p$-form $B$ to a $p-1$-form $^\star{B}\equiv{^*d\,^*}B$.
For example, acting on the Faraday 2-form $F$, one finds ${^\star{F}}=\nabla^\mu F_{\mn}dx^\nu$.
With this notation, both the BP and LW operators can be derived from the same classical action 
$-\frac{\lambda}{2\Lambda^2}\int(^\star{F})\wedge(^*{^\star{F}})$, up to a boundary term.
However, this equivalence does not survive in curved spacetime. 
The covariant LW and BP operators,
$\frac{1}{4}\nabla_\mu F_{\alpha\beta}\nabla^{\mu}F^{\alpha\beta}$
and $\hf\nabla_\mu F^{\mu\nu}\nabla^{\rho}F_{\rho\nu}$,
differ by non-minimal photon-curvature couplings, 
as follows from Eqs. (12-13) of Ref. \cite{Antisymm}.

The only dimension-6 Lorentz- and gauge-invariant photon operator \cite{Schreck18} is 
\bea&&\label{LI-d6}
I_\mathrm{LI}\subset\int\left[-\frac{\lambda}{2\Lambda^2}{(^\star{F})\wedge(^*{^\star{F}})}
\right],
\eea
where $\Lambda$ denotes an unknown ultraviolet scale and $\lambda$ is a dimensionless coupling constant.
First, $dF$ cannot appear in the Lagrangian since $F=dA$ and thus $dF\equiv0$.
Second, one may naively consider quadratic operators such as
$\frac{\lambda}{\Lambda^2}d{^*F}\wedge{^\star{F}}$ or $\frac{\lambda}{\Lambda^2}{d(^\star F)}\wedge{^*{F}}$.
However, these are equivalent to the BP or LW operator in Eq. (\ref{LI-d6}),
up to a total derivative and irrelevant numerical factors. 
Indeed, one may verify the identity
$(^\star{F})\wedge(^*{^\star{F}})=d{^*F}\wedge{^\star{F}}$.
Third, we may naively introduce a dimension-6 operator of the form
$\frac{\zeta}{M^2}d\theta\wedge{^\star{F}}\wedge F$,
resembling an axion-photon coupling.
However, the induced correction to the equations of motion 
takes the form
$\frac{\zeta}{M^2}[\prt_\mu\theta\,\prt^\nu-\prt_{\mu}^{~\nu}\theta]\prt_\rho F^{\rho\mu}$. 
Since this term vanishes upon imposing the leading-order Maxwell equation
$\prt_\rho F^{\rho\mu}=0$,
the operator is redundant at leading order and affects physical observables only at higher orders.

The lesson from the higher-derivative operators above is that,
in general, we should not allow wedge products involving either $^\star F$ or $d^*F$,
since they may generate equation of motion reducible operators, such as
$d\theta\wedge{^\star{F}}\wedge F$ or $i_\xi F\wedge\mathcal{L}_\xi d^*F$. 
In this sense, the LW operator in Eq. (\ref{LI-d6}) is exceptional.
Allowing background tensor–photon couplings enlarges the possible set of dimension-6 operators, 
albeit at the cost of Lorentz invariance. 
A representative class of operators constructed from a background vector
$\xi^\mu$ is
\bea&&\label{LV-d6}
I_\mathrm{LV}\subset
\int\left[\rho(i_\xi{^\star F})^2{^*1}+\gamma(i_\xi d^*F)\wedge{^*}(i_\xi d^*F)\right]\nn&&
+\int\mathcal{L}_\xi({^*i}_\xi F)\wedge [\mathcal{L}_\xi (\alpha\,i_\xi F+\beta\,i_\xi {^*F})],
\eea
where $\alpha,~\beta,~\gamma,~\rho$ are coupling constants.
When evaluated in flat spacetime or in Fermi normal coordinates (FNC), 
the first term resembles the Bopp–Podolsky operator, 
whereas the remaining terms are analogous to the Lee–Wick operator. 
Taken together, they may be regarded as a dimension-6 generalization of the Myers–Pospelov operator.

Interestingly, upon identifying $D_{\alpha\beta}=\xi_\alpha\xi_\beta$, 
the first term 
$\int\rho(i_\xi{^\star F})^2{^*1}=\int\sqrt{-g}\,d^4x\rho\,\xi^\alpha\xi^\beta\prt_\mu F^{\mu}_{~\alpha}\prt_\nu F^{\nu}_{~\beta}$
coincides with the dimension-6 anisotropic operator proposed in Ref. \cite{Schreck18}.
Unlike the other operators, however, it is not a 4-form built purely from wedge products of dynamical fields; 
instead, it is the product of a scalar (a 0-form obtained by contracting the 1-form $^\star F$) 
with the invariant volume form $^*1$.
The propagator associated with the term $D_{\alpha\beta}\nabla_\mu F^{\mu}_{~\alpha}\nabla_\nu F^{\nu}_{~\beta}$ 
was derived analytically in Ref. \cite{Schreck18}, 
where the dispersion relation, unitarity, and causality were analyzed, 
and a viable parameter space compatible with quantization and free of acausal behavior was identified.

As for the operators in the last line, they involve quadratic Lie derivatives 
and are included merely to illustrate that, in principle, 
one may always construct higher-order operators with increasing numbers of Lie derivatives, 
provided the mass dimensions are properly matched. 
For example, the operators in Eq. (\ref{LV-d6}) are of mass dimension six, 
where we implicitly assume $\mathrm{dim}[\xi^\mu]=0$.
One might expect that allowing more general background tensor structures would generate additional dimension-6 LV photon operators. 
However, experimental constraints \cite{dataTables} indicate that LV background tensor coefficients are extremely small, 
so only operators linear in these coefficients should be retained. 
The appearance of higher powers of $\xi$ in Eq. (\ref{LV-d6}) reflects the fact 
that they can be identified with products of a given tensor, \eg,
$D_{\alpha\beta}=\xi_\alpha\xi_\beta$.

In four-dimensional spacetime, the highest nontrivial differential form is degree four. 
Introducing a background 1-form $b=b_\mu\,dx^\mu$ or a 2-form $H=\hf H_{\mn}dx^\mu\wedge dx^\nu$ does not modify photon kinematics,
since the only gauge-invariant 2-forms available for wedge products are $F$ and $^*F$.
Therefore, without introducing a vector for contraction or a Lie derivative, 
one cannot construct a non-vanishing gauge-invariant 4-form that alters standard Maxwell dynamics.

As in the dimension-5 Myers–Pospelov case, 
the dimension-6 extended Maxwell operators also suffer from stability issues 
typical of higher-derivative theories. 
For instance, the Podolsky theory contains two modes: 
the usual massless photon mode and an additional massive mode. 
In momentum space, the latter has a wrong-sign residue 
and is therefore interpreted as a ghost in canonical quantization. 
Alternatively, it may be viewed as a Lee–Wick mode, 
which improves renormalizability at the cost of mild microcausality violation. 
As noted above, similar issues arise for dimension-6 LV photon operators \cite{Schreck18, Marco2023}. 
These pathologies, however, do not undermine the theory when viewed as an effective field theory (EFT). 
The theory, including high-derivative operators, 
is not intended as a fundamental description but rather as a low-energy extension valid below a certain cutoff scale $\Lambda$. 
Within the EFT framework, operators are organized as an expansion in powers of $1/\Lambda$, 
with higher-order terms increasingly suppressed. 
Truncating the expansion at a fixed mass dimension 
and treating the resulting corrections perturbatively yields reliable equations of motion for low-energy phenomena, 
including leading LV effects. 
Ghost modes and other instabilities -- such as the massive mode in Podolsky theory -- 
lie outside the regime of validity of the EFT. 
They merely indicate where the effective description breaks down and where ultraviolet completion must take over.

\subsection{The extended operator construction -- a short summary}

To construct a quadratic Maxwell Lagrangian that preserves gauge invariance
while allowing Lorentz and CPT violation, we consider the following ingredients: 
the gauge potential $A=A_\mu dx^\mu$, the metric tensor $g_{\mn}$, 
and a set of background vectors $\{\xi_i=\xi_i^\mu\prt_\mu,~i=1,...,n\}$.
Two points are worth emphasizing:
1. restricting the LV extension to terms quadratic in $A$ (or equivalently in $F$) 
is necessary to maintain the linearity of the modified Maxwell equations; 
higher-order terms would instead modify photon vertices rather than the propagator.
Since experimental tests of Lorentz and CPT invariance primarily probe kinematic effects, 
propagator corrections provide the leading signals, and higher-order terms can be neglected at this level.
2. We treat the four-potential 1-form $A=A_\mu dx^\mu$ as the fundamental field 
and define the Faraday 2-form off shell as $F=dA$,
so that the Bianchi identity $dF=0$ holds identically.
Accordingly, we exclude Palatini-like formulations 
in which $F$ is varied independently and the Bianchi identity is enforced dynamically 
(e.g., via a term $S\supset\int[A\wedge dF]$).
This parallels the distinction between metric and Palatini formulations in gravity, where
the metric $g_{\mn}$ and the affine connection $\Gamma^\rho_{\mn}$ are treated as independent variables.

With these ingredients in hand, we can construct the following differential forms:
\bit
\item 1-forms: 
$A=A_\mu dx^\mu$ and $^\star{F}$.
Given a background vector $\xi$, we can form the 1-forms
$i_{\xi}F$, $i_{\xi}{^*F}$ and $\tilde{\xi}=g(\xi,\cdot)$,
where $g$ is the metric tensor. 
Here the tilde on $\xi$ indicates the 1-form obtained by lowering the index with the metric,
\ie, $\tilde{\xi}\cdot\zeta=g(\xi,\zeta)$ for any vector $\zeta$.
\item 2-forms: 
The fundamental gauge-invariant 2-forms are $F=dA$ and its Hodge dual $^*F$.
Additional 2-forms can be constructed, such as
$d{^\star}F=(d^*)^2F$ and $i_{\xi}{d^*F}$, and the Lie derivatives $\mathcal{L}_\xi F$, $\mathcal{L}_\xi{^*F}$.
\item 3-forms:  
$^*A$ and $d^*F$. Given a background vector $\xi$,
we may also form $^*(i_{\xi} F)$ and $^*(i_{\xi}{^*F})$.
\item 4-forms: 
the fundamental 4-form is the invariant volume form $*1$.
The basic Lorentz-invariant 4-forms include $F\wedge F$ and $F\wedge{^* F}$.
\eit
To construct photon operators beyond the conventional Lorentz-invariant terms such as
$F\wedge F,~F\wedge{^*F}$, 
we may take wedge products of the above building blocks to form gauge-invariant 4-forms.
For example, $\mathcal{L}_\xi(i_\xi ^*F)\wedge(^*i_\xi F)$ reproduces the operator 
in Eq. (\ref{DfMP}) upon using identity (\ref{doublep-form}).
For systematic construction, it is convenient to classify the building blocks by form degree and mass dimension, 
and further distinguish them according to their dependence on background tensor coefficients. 
Operators marked with ``$^@$" in the upper right denote independence from the background vector $\xi$. 
For simplicity, we list only the case of a vector background field. 
These building blocks are summarized in Table \ref{BDF}.

Note that we allow multiple appearances of background vectors by identifying them as 
components of a single tensor,
\eg, $D_{\alpha\beta}=\xi_\alpha\xi_\beta$ and $(c_F)_{\mn}\propto\zeta_{(\mu}\xi_{\nu)}$,
This does not mean that we include nonlinear LV coefficients in our construction.
Two further points are worth emphasizing.
\bit
\item 1. We intentionally reserve a place for the volume form $^*1$ as a special 4-form. 
In principle, this allows one to construct photon operators by multiplying $^*1$
with an observer Lorentz scalar, such as $(i_\xi^\star F)^2$ in Eq. (64). 
However, such terms effectively decouple the photon dynamics from the basic differential-form $[\mu\nu\rho\sigma]$. 
As a result, our differential-form approach for translating the extended Maxwell equations into the Newman–Penrose formalism 
does not apply to this class of operators. 
We nevertheless keep it to emphasize that not all admissible operators can be written as essential wedge products of dynamical forms; 
some may instead appear as scalars multiplying $^*1$.
\item 2. Another important guiding principle is to avoid operators linear in $d^*F$,
since they may be reducible by the equations of motion. 
By contrast, operators quadratic in $d^*F$, such as  $i_\xi d^*F\wedge^*(i_\xi d^*F)$ in Eq. (\ref{LV-d6}), are allowed: 
they modify the dispersion relation and cannot be removed by field redefinitions 
or by using the leading-order vacuum Maxwell equation $\prt_\mu F^{\mn}=0$.
\eit

We also emphasize that our procedure generates only a subset of the LV photon operators, 
corresponding to the last five rows of Table III in Ref. \cite{SMEgauge}.
In Ref. \cite{SMEgauge}, LV photon operators are presented as part of the broader class of Abelian gauge operators. 
Table III explicitly lists photon operators up to mass dimension six 
and additionally provides explicit examples of several operators up to dimension eight.
Many higher-dimensional operators there are nonlinear in the field strength $F_{\mn}$
and therefore lie beyond the quadratic framework considered here.
Note we also omit the tadpole operator linear in $A_\mu$.
While such a term is allowed from the EFT perspective, 
it lies beyond the quadratic class we interested in this work.
For nonlinear extensions of electrodynamics, the interested reader may also resort to
\cite{NonLinEM1, NonLinEM2}.

With additional assumptions, such as introducing linear maps $\kappa:\Lambda^2(M)\mapsto\Lambda^2(M)$ or $\chi$, 
a broader class of operators can be recast in terms of differential forms, 
in a manner analogous to Eqs. (\ref{k5decomp}) and (\ref{k6decomp}). 
Although decomposing an observer background tensor into tensor direct products
significantly reduces the number of independent degrees of freedom, 
at least a subset of the $k^{(5)}$ and $k^{(6)}_\partial$ operators admits a compact differential-form representation.
This perspective may also suggest a systematic way to express more general operators in the language of differential forms,
provided that additional structures, such as linear or multi-linear map, is introduced. 
For instance, by defining a multi-linear map $K:\Lambda^2(M)\otimes\Lambda^2(M)\mapsto\Lambda^2(M)$, 
which maps two 2-forms to a 2-form, 
certain nonlinear operators can likewise be formulated within this framework. 
Given two 2-forms $F$ and $G$, such a map may be written as
\bea\label{multilinMap}
K(F,G)=\frac{1}{2^4}K_{\mn}^{~~\alpha\beta\rho\sigma}F_{\alpha\beta}G_{\rho\sigma}[\mn].
\eea
In this manner, nonlinear operators such as $k^{(6)}_F$ and $k^{(8)}_F$ can be written compactly in differential forms.
For example, $\int\sqrt{-g}d^4x\, k^{(6)\alpha\beta\mn\rho\sigma}_F F_{\alpha\beta}F_{\mn}F_{\rho\sigma}=\int(^*F)\wedge K(F, F)$,
provided one identifies $k^{(6)\alpha\beta\mn\rho\sigma}_F=-\frac{3}{4}K^{\mn\alpha\beta\rho\sigma}$.

Throughout this work, we assume that the background tensors originate from spontaneous Lorentz symmetry breaking. 
Under this assumption, it is unnecessary to distinguish coefficients with upper and lower indices
(\eg, $k^{\mn...}$ versus $k_{\mn...}$), otherwise, a more intricate analysis would be required \cite{LVgrav}. 
In a genuine curved spacetime, additional curvature-photon couplings permit a much broader class of extended LV operators,
see, Table XIV of Ref. \cite{LVgrav}. 
However, such constructions are not restricted to purely quadratic photon operators, 
and are not built solely from fundamental geometric operations such as exterior differentiation, the Hodge dual, 
and Lie derivatives. 
They therefore lie beyond the scope of our present work and are deferred to future investigation. 

\begin{widetext}
\begin{table*}[ht]
\renewcommand{\arraystretch}{1.5}
    \centering
    \begin{tabular}{|c|c|c|c|c|c|}
	\hline
	\diagbox{p-form}{mass dimension} &  $d=1$  &  $d=2$                      &  $d=3$         & $d=0/1$  \\
	\hline
	$p=1$  & ${A}^@$ &  $i_{\xi} F$,~$i_{\xi}{^*F}$ & ${^\star F}^@$, $\mathcal{L}_\xi(i_\xi ^*F)$,
    $\mathcal{L}_\xi(i_\xi F)$,...& $\tilde{\xi}=g(\xi,\cdot)$   \\
	\hline
	$p=2$  & $\tilde{\xi}\wedge A$ & ${F}^@$,~${^*F}^@$ & $d{^\star F}^@$,~$i_\xi d^*F$,~$\mathcal{L}_\xi F$,~$\mathcal{L}_\xi({^*F})$ &  \\
	\hline
	$p=3$  & ${^*A}^@$ & $^*i_{\xi} F$,~$^*i_{\xi}({^*F})$ & ${d^*F}^@$,  $^*\mathcal{L}_\xi(i_\xi ^*F)$, $^*\mathcal{L}_\xi(i_\xi F)$ &  $^*\tilde{\xi}$ \\
	\hline
	\hline
\end{tabular}
  \caption{Table of some building blocks of photon operators. The operators with superscript $@$ on the upper right mean they are essentially dynamical, and operators building pure of these operators, say $d^*F\wedge{^\star F}$, are Lorentz invariant. All the remain operators are background vector dependent and therefore cannot maintain Lorentz invariance in general. }
    \label{BDF}
\end{table*}
\end{widetext}



\section{Summary}\label{summ}
In this article, we develop a differential-form approach (DFA) to derive the Newman–Penrose (NP) formulation of the Maxwell equations, 
following the strategy of Ref. \cite{ZX2025}.
This approach exploits the coordinate independence of differential forms 
and may be viewed as complementary to the intrinsic-derivative method. 
Such coordinate independence is a fundamental consistency requirement of any viable physical theory and, 
in the context of Lorentz-violating (LV) frameworks, corresponds to observer diffeomorphism invariance \cite{LVgrav}.
Within the DFA, both the extended Maxwell actions and their field equations admit compact representations, 
and the NP formulation of the extended Maxwell equations can be obtained systematically. 
Using this framework, we have explicitly derived the NP-form Maxwell equations for power-counting renormalizable LV operators 
in Secs. III A and III B; see Eqs. (\ref{CPTO-NP-Maxwell}) and (\ref{CPT-E-cNP}).
Although Eq. (\ref{CPT-E-cNP}) includes only a subset of the dimension-four LV background coefficients,
namely, the Ricci-like sector of $\kappa_{\mn\alpha\beta}$ [see Eq. (\ref{PO-CPTE-photon})], 
the full set of  $\kappa$-corrected Maxwell equations in NP form can be obtained straightforwardly 
by applying the same procedure once the differential-form equation (\ref{kFDiff}) is specified. 
We do not present these expressions explicitly, as they are algebraically lengthy and offer no new conceptual insight.

Beyond infrared-relevant or marginal corrections to electrodynamics, 
we also consider dimension-five and dimension-six operators. 
These operators are strictly irrelevant in the infrared, 
and consequently the NP formalism -- being intrinsically adapted to null-propagating infrared degrees of freedom 
such as photons and gravitons -- 
does not play an essential role in their analysis. 
We nevertheless include them for completeness. 
More importantly, by explicitly constructing both the action and the corresponding extended Maxwell equations, 
we illustrate the compactness and organizational clarity afforded by the differential-form formalism. 
Furthermore, in constructing both the LI and LV photon operators (see Secs. \ref{d5HDE} and \ref{d6HDE}),
we show that, at least up to mass dimension six, 
the building principles remain systematic and transparent: 
the operators can be classified by counting independent differential forms while eliminating redundancies.

To obtain the above equations in a sufficiently simple form, 
we have adopted the assumption that 
the covariant derivatives of the background tensor fields vanish in a special observer frame, 
namely Fermi normal coordinates (FNC). 
For example, for a background vector $\xi^\mu$, this implies $\nabla_\mu\xi^\nu=0$ in FNC.
Physically, such a frame may be identified with the Sun’s free-fall frame relative to the Galactic center or, 
more generally, with an approximately inertial frame in a weak-gravity environment, 
such as that of an observatory.
Importantly, {\it the DFA used to derive the Maxwell equations in NP form does not rely on this assumption}; 
it is introduced solely to simplify the calculations. 
Without it, additional terms involving intrinsic derivatives of background-tensor projections onto the null tetrad, 
such as $\xi^a_{~|b}$, would appear.

In principle, this formalism can be refined by adopting a quasi-null tetrad adapted to an effective metric, 
since in LV theories photons generally do not propagate along null geodesics defined by the conventional metric $g_{\mn}$, 
assuming the test-particle limit in which gravity (and hence the spacetime geometry) remains intact. 
Nevertheless, because the standard LI null tetrad $\{E_a^{~\nu}\prt_\nu,~a=1,...,4\}=\{\bf{l},\bf{n},\bf{m},\bf{\bar{m}}\}$ forms a complete basis, any alternative tetrad --- including a quasi-null one --- can always be expanded in terms of $E_a^{~\nu},~a=1,...,4$
[see Eq. (\ref{completeN}) and (\ref{QuasiNull})].
For the purpose of illustrating how the DFA facilitates the derivation of Maxwell equations in NP form, 
the standard null tetrad is therefore sufficient and may be regarded as a zeroth-order approximation. 
Developing a more systematic approximation scheme based on quasi-null tetrads 
could nevertheless be worthwhile for analyzing the asymptotic behavior of LV photons and gravitons, fully exploiting the NP formalism.

More broadly, upon introducing additional structures such as linear and multi-linear maps
(see Eqs. (\ref{k5decomp}), (\ref{k6decomp}), and (\ref{multilinMap}),
a substantially wider class of photon operators, including those beyond purely quadratic photon couplings,
can be formulated in terms of differential forms. 
Such operators in the Lagrangian density are catalogued, for example, 
in Table III of Ref. \cite{SMEgauge} and Table XIV of Ref. \cite{LVgrav},
and merit further investigation in their own right.

Despite this broader landscape, 
the differential-form framework developed here provides a versatile and systematic tool 
for deriving extended Maxwell equations, whether for studying asymptotic behaviors \cite{ZX2025} 
or for analyzing polarization evolution. 
The latter will be investigated in future work. 
In either context, the combination of differential forms with the null formalism offers a coherent and powerful approach
to investigate the impact of background tensor fields on photon propagation. 

\section*{Acknowledgements}
We express our gratitude to A. Kosteleck\'y for his valuable comments and suggestions.
Tao Zhu is supported by the National Natural Science Foundation of China under Grant No.~12275238, No.~12542053, and No.~11675143, the National Key Research and Development Program of China under Grant No. 2020YFC2201503, the Zhejiang Provincial Natural Science Foundation of China under Grants No.~LR21A050001 and No.~LY20A050002, and the Fundamental Research Funds for the Provincial Universities of Zhejiang in China under Grant No.~RF-A2019015.

\appendix

\section*{Appendix A: Standard procedure in establishing null tetrad}\label{Procedure-NT}  

The general procedure (though it may differ from more personalized approaches with individual experience)
for constructing a null tetrad adapted to a given spacetime metric is as follows:\\
a. First, we may find out {\it the hypersurfaces of light wavefront parameterized by $u=\mathrm{const.}$},
which are intrinsically light-like.
With a specific $u=\mathrm{const.}$, we can define the first covariant null vector as
\bea\label{Nulltrad-1st}
l_\mu\equiv-\nabla_\mu{u},
\eea
which is orthogonal to the hypersurface $u=\mathrm{const.}$, since $l_\mu$ represents component of the 1-form ${\bf{l}}=l_\mu dx^\mu=-du$.
The null hypersurface $u=\mathrm{const.}$ also indicates that
$l_\mu$ is both orthogonal and tangent to it.
The contravariant null vector is $l^\mu=-g^{\mn}\nabla_\nu{u}$.
According to the torsion-free condition,
we can further confirm that the null curve, to which $l_\mu$ is tangent to, is a null geodesic,
because
\bea&&
\nabla_l l^\mu=g^{\al\be}\nabla_\be{u}\,\nabla_\al(g^{\mn}\nabla_\nu{u})
=g^{\al\be}\nabla_\be{u}\,g^{\mn}\nabla_\al(\nabla_\nu{u})\nn&&
=g^{\al\be}\nabla_\be{u}\,g^{\mn}\nabla_\nu\nabla_\al{u}
=\frac{g^{\mn}}{2}\nabla_\nu(g^{\al\be}\nabla_\be{u}\nabla_\al{u})\nn&&
=0,
\eea
provided $\nabla_\nu{g}^{\al\be}=0$ and $\Gamma^\rho_{[\al\be]}=0$.
The last equation holds because $g^{\al\be}l_\al l_\be=0$, \ie, $l_\al$ is null.

Now we can directly chose $u$ as the first coordinate
\bea\label{Nullcoord-1st}
l_\mu=-\nabla_\mu{u}=-\delta^0_\mu.
\eea
The reason $u$ is chosen as the first coordinate is that
{\it $u$ naturally serves as a time parameter for radiation.
As the light wave propagates, the constant-$u$ hypersurfaces representing
equal phase wavefronts also evolve along with it.}
Thus the normal direction given by $-{\bf{l}}=du$ indicates the direction of increasing time.

b. Subsequently, we may choose the affine parameter $r$ of the null geodesics lying on the $u=\mathrm{const.}$
hypersurface as the second natural coordinate.
Thus, we have
\bea\label{Nulltrad-2nd}&&
l^\mu=\frac{dx^\mu}{dr}=g^{\mn}l_\nu=-g^{\mn}\nabla_\nu{u}=-g^{\mu0}\nn&&
\Rightarrow g^{\mu0}=-\delta^\mu_1~(g^{01}=-1),~~\hat{l}=l^\mu\prt_\mu=\prt_r
\eea
where we have used Eqs. (\ref{Nulltrad-1st}) and (\ref{Nullcoord-1st}).
It is easy to check the null condition of the scalar product $(\hat{l},{\bf{l}})=(\prt_r,-du)=0$.
Then we define the second null vector $n^\mu$ to satisfy the quasi-normal condition $(\hat{l},{\bf{n}})=(\hat{n},{\bf{l}})=n^\mu l_\mu=-1$.
This condition allow us to determine either the covariant or contravariant vector as following,
\bea
{\bf{n}}=-dr+Vdu+Y_Adx^A,\quad \hat{n}=\prt_u+U\prt_r+X^A\prt_A,
\eea
where $U,V,X^A,Y_A\in\mathbb{R}$ and $A=2,3$.
From the fact that $n^\al=g^{\al\be}n_\be$, we can establish the relation between these coefficients as below
\bea&&
\hspace{-5mm}n^\al=g^{\al0}V-g^{\al1}+g^{\al A}Y_A=\delta^\al_0+U\delta^\al_1+X^A\delta^\al_A\nn&&
\hspace{-5mm}\Rightarrow -V-g^{11}+g^{1A}Y_A=U,~~ g^{AB}Y_A-g^{B1}=X^B.
\eea

c. For the remain two null vectors, since they are orthogonal to both $\hat{n},~\hat{l}$,
and given that any hyperbolic Riemannian space admits only two independent intrinsic null vectors,
{\it the remain two orthogonal vectors must be spacelike
and they span the transversal space orthogonal to the direction of wave propagation}
--- precisely the directions denoted by $l^\mu,~n^\mu$ in 4-dimensional spacetime.
By allowing complex extension, we can construct, from a pair of spacelike basis vectors
$\xi^\mu,~\zeta^\mu$ lying in the transversal space,
the null vectors $\frac{1}{\sqrt{2}}(\xi^\mu\pm i\zeta^\mu)$,
which will be denoted later by $\hat{m},~\hat{\bar{m}}$.
Since the basis vectors $\xi^\mu,~\zeta^\mu$ are intrinsically spacelike, we have
\bea\label{Nulltrad-3rd}
\hat{m}=\omega\,\prt_r+\xi^A\,\prt_A, \quad \hat{\bar{m}}=\bar{\hat{m}},
\eea
where the overbar on the head of $\hat{m}$ means complex conjugate
and thus $\omega,~\xi^A\in\mathbb{C}$ in general.

Note that the degrees of freedom (d.o.f.) must match in all equivalent formulations.
Firstly, the above construction leads us to the Bondi metric
{\small
\bea\label{BondiMetrics}
(g^{\mn})=\left(
          \begin{array}{cccc}
            0 & -1 & 0 & 0 \\
            -1 & g^{11} & g^{12} & g^{13} \\
            0 & g^{12} & g^{22} & g^{23} \\
            0 & g^{13} & g^{23} & g^{33} \\
          \end{array}
        \right)
        \Leftrightarrow
(g_{\mn})=\left(
            \begin{array}{cccc}
              g_{00} & -1 & g_{02} & g_{03} \\
              -1 & 0 & 0 & 0 \\
              g_{02} & 0 & g_{22} & g_{23} \\
              g_{03} & 0 & g_{23} & g_{33} \\
            \end{array}
          \right).\nonumber
\eea
}
\hspace{-0.5mm}It is clear that there are only $6$ independent metric components.
With the completeness relation (\ref{completeN}),
we would expect the same number of parameters in the null tetrad.
However, a naive counting of the real parameters $U,X^A$ and complex parameters $\omega,~\xi^A$
yields the number $9$, which is greater than $6$.
However, further investigation reveals that
{\it there is some arbitrariness in defining null tetrad}:

1. there is a $U(1)$ rotational d.o.f. associated with the definition of the two complex null vectors,
\bea\label{NullTransf-1}
m^\nu\equiv\frac{1}{\sqrt{2}}[\xi^\nu+i\zeta^\nu]
\rightarrow m^\nu\,e^{i\phi},\quad \bar{m}^\nu\rightarrow \bar{m}^\nu\,e^{-i\phi}.
\eea
This $U(1)$ rotational d.o.f. also allows us to define the spin weight for a give function.
For example, a rotational scalar function $f=f_\mu m^\mu$ has spin weight $+1$,
while its complex conjugate $\bar{f}=\bar{f}_\mu\bar{m}^\mu$ has spin weight $-1$.

2. there is a transformation keeping the direction $l^\mu$ unchanged,
\bea&&\label{NullTransf-2}
l^\mu\rightarrow l^\mu, \quad m^\mu\rightarrow m^\mu+B\,l^\mu,\nn&&
n^\mu\rightarrow n^\mu+\bar{B}\,m^\mu+B\,\bar{m}^\mu+B\bar{B}\,l^\mu.
\eea
To make $l^\mu$ unaltered gives the first transformation $l^\mu\rightarrow l^\mu$ in (\ref{NullTransf-2}),
while to make $m^2=0,~m\cdot l=0$, $m^\mu$ can be changed only by adding a factor proportional to $l^\mu$.
Since $m^\mu$ is a complex null vector, the proportional constant may also be complex, \ie, $B\in\mathbb{C}$.
The last transformation is then dictated by the requirements that $n\cdot l=-1,~n\cdot m=0,~n^2=0$ and $n^\mu$ is a real null vector.

Counting the number of real transformation parameters $B\in\mathbb{C},~\phi\in\mathbb{R}$ gives a total of 3.
Thus the number of independent degrees of freedom is $9-3=6$,
exactly matching the number of independent components of the Bondi metric.
With the upper indices version of the completeness relation (\ref{completeN}),
we can express the metric components in terms of the known parameters of the null tetrad,
\bea&&
g^{11}=2(\omega\,\bar{\omega}-U),\quad
g^{1A}=(\xi^A\bar{\omega}+c.c.)-X^A,\nn&&
g^{AB}=\xi^A\bar{\xi}^B+c.c.,\quad g^{01}=g^{10}=-1,
\eea
where $A,~B=2,3$.
Note, we have used the notation of Refs. \cite{NP1962-68, Chand-MToBH} but with signature $+2$ instead of $-2$,
so some of our results differ by a total minus sign from those in Refs. \cite{NP1962-68, Chand-MToBH}.

\section*{Appendix B: Some identities of differential forms}
In this appendix, we collect several identities of differential forms relevant to this work. 
\begin{widetext}
For brevity, most are given without proof, 
since they are elementary results available in standard references on differential forms.
Suppose $\alpha,~\beta$ are two p-forms, then we have
\bea&&\label{doublep-form}
\alpha\wedge(^*\beta)=\frac{1}{(p!)^2\,q!}\alpha_{i_1...i_p}\epsilon_{j_1...j_q}^{~~~~k_1...k_p}
\beta_{k_1...k_p}[dx^{i_1}\wedge...\wedge dx^{i_p}]\wedge[dx^{j_1}\wedge...\wedge dx^{j_q}]\nn&&
=\frac{(-1)^{p\,q}}{p!}\alpha_{i_1...i_p}\beta^{i_1...i_p}\sqrt{|g|}[dx^1\wedge...\wedge dx^n]
=\beta\wedge(^*\alpha),
\eea
where $q=n-p$ and $\mathrm{dim}\mathcal{M}=n$, and $\#t$ is the number of time directions.
The second-to-last line is because
$\epsilon^{i_1...i_n}\equiv\frac{(-1)^{\#t}}{\sqrt{-g}}[i_1...i_n]$, where $[i_1...i_n]$
is the Levi-Civita symbol,
and thus we arrive at $\epsilon_{j_1...j_qk_1...k_p}
\epsilon^{j_1...j_qi_1...i_p}=(-1)^{\#t}\delta^{i_1...i_p}_{k_1...k_p}p!q!$.
So, whatever the definitions, we will have the wedge product identity
$\alpha\wedge(^*\beta)=\beta\wedge(^*\alpha)$.
In fact, the sign factor $(-1)^{pq}$ can be eliminated if we verify $(^*\alpha)\wedge\beta=(^*\beta)\wedge\alpha$ instead.
\end{widetext}

Suppose there is a vector $\xi=\xi^\mu\prt_\mu$, and a p-form $\alpha$ and $r$-form $\gamma$, then we have the following graded Leibnitz rules
for the contraction and exterior differential operators $i_\xi,~d$, and the usual Leibnitz rule for Lie derivatives $\mathcal{L}_\xi$,
\bea&&
d[\alpha\wedge\gamma]=
(d\alpha)\wedge\gamma+(-1)^p\alpha\wedge(d\gamma),\\&&
i_\xi[\alpha\wedge\gamma]=
(i_\xi\alpha)\wedge\gamma+(-1)^p\alpha\wedge(i_\xi\gamma),\\&&
\mathcal{L}_\xi[\alpha\wedge\gamma]=(\mathcal{L}_\xi\alpha)\wedge\gamma+\alpha\wedge(\mathcal{L}_\xi\gamma).
\eea

Also there are several identities for
\bea&&
d^2\alpha=0,\quad i_n^2\alpha=0,\quad
\alpha\wedge\gamma\equiv0.
\eea
where the first two equations is due to the anti-symmetry of the definition of a p-form $\alpha$,
while the third equation is valid provided $p+r\ge n+1$.
Note the second equation together with Cartan's magic formula implies another identity
\bea&&
\mathcal{L}_\xi\, i_\xi-i_\xi\, \mathcal{L}_\xi
=d\,i_\xi^2-i_\xi^2\,d=0\Rightarrow
[\mathcal{L}_\xi,\, i_\xi]=0.
\eea
This commutativity in fact
comes from a more general identity of commutator equation
\bea
[\mathcal{L}_\xi,\, i_\zeta]\alpha
=i_{[\xi,\,\zeta]}\alpha,
\eea
when $\xi=\zeta$, $i_0\alpha=0$ and hence the commutativity $[\mathcal{L}_\xi,\, i_\xi]=0$.
Aside from the commutativity between Lie derivative and the contraction along the same vector, there is also a commutativity
\bea
[\mathcal{L}_\xi,\,d]=0,
\eea
which can be quite easy to verify by artan¡¯s magic formula.
There are also some identities we need to verify
\bea&&
i_\xi\alpha=(-1)^{q+(p-1)(q+1)}{^*[\xi^\flat\wedge(^*\alpha)]},
\\&&
^*(i_\xi\alpha)=(-1)^{q+t}\xi^\flat\wedge(^*\alpha)
\eea
where $\xi^\flat=\xi_\mu dx^\mu$ is the dual 1-form for a given vector $\xi=\xi^\mu\prt_\mu$ if metric is assigned to the manifold $\mathcal{M}$, and $t$ is the number of time-like directions (in usual cases, $t=1$).

\section*{Appendix C: Maxwell action and definition of $k_F$ dual}
Firstly, we define the dual of $k_F$ (denoted as $\kappa$ for simplicity) in analogy to the Hodge dual. In fact, as a map between 2-forms, $\kappa$ can be viewed as a mixed
tensor belonging to $\Lambda^2_2(M)$, and thus one possible way is to decompose $\kappa$ as a direct product of a bivector $\xi^{[c}\zeta^{d]}$ composed of two vectors and a 2-form $\omega$, in other words,
\bea&&
\kappa_{ab}^{~cd}=\hf\left(\zeta^{[c}\xi^{d]}\omega_{ab}-\zeta_{[a}\xi_{b]}\omega^{cd}\right),\\&&
\omega_{ab}=-\omega_{ba},\quad
\omega^{ab}=g^{ac}g^{bd}\omega_{cd},\nn&&
\xi_c=g_{cd}\xi^d,\quad \zeta_c=g_{cd}\zeta^d,
\eea
where $\zeta\equiv\zeta^a e_a,~\xi\equiv\xi^a e_a$ are two vectors
and $\omega\equiv\hf\omega_{cd}\,e^c\wedge{e^d}$ is a 2-form,
and $\zeta_a,~\xi_b$ and $\omega^{ab}$ are the covariant and contravariant components lowered and raised by the metric tensors $g_{cd}$ and $g_{ab}$, respectively.
Or it is possible to decompose $\kappa$ as
\bea&&
\kappa_{ab}^{~~cd}=\hf[(c_F)_a^{~c}\delta_b^d-(c_F)_a^{~d}\delta_b^c-(c_F)_b^{~c}\delta_a^d+(c_F)_b^d\delta_a^{~c}],\nn&&
(c_F)_a^{~b}\equiv\hf[\xi_a\zeta^b+\xi_b\zeta_a].
\eea
Since we may regard $\kappa\in\Lambda^2_2(M)$ as a linear map
between 2-forms, we may define a dual 2-form given a Faraday
2-form $F$:
\bea
{^{\kappa}{F}}\equiv\hf{^{\kappa}{F}}_{ab}\,e^a\wedge{e^b}=\frac{1}{2^2}\kappa_{ab}^{~~cd}F_{cd}\,e^a\wedge{e^b}.
\eea

Now we give a short review of the Maxwell action written in differential forms:
\bea&&
-\hf\int F\wedge{^*F}
=-\frac{1}{2^4}\int{\epsilon}_{cd}^{~~mn}F_{mn}F_{ab}e^a\wedge{e^b}\wedge{e^c}\wedge{e^d}\nn&&
=\frac{-1}{4}\int\sqrt{-g}\,d^4x\,F^{ab}F_{ab},
\eea
where we used the definition of dimensional four invariant volume
$\sqrt{-g}[dx^0\wedge...\wedge dx^3]=\frac{\epsilon_{abcd}}{4!}dx^a\wedge...\wedge dx^d$ and the identity $\epsilon_{cdmn}\epsilon^{cdab}=-2!\delta^{ab}_{mn}=2[\delta^a_n\delta^b_m-\delta^a_m\delta^b_n]$.

Similarly, the LV CPT-even action reads
\bea&&
\int(^*F)\wedge(^{\kappa} F)=\frac{1}{2^4}\int
\epsilon_{ab}^{~~mn}F_{mn}
e^a\wedge{e^b}\wedge
\kappa_{cd}^{~~pq}{F}_{pq}e^c\wedge{e^d}\nn&&
=\frac{1}{2^2}\int\sqrt{-g}d^4x\,
F^{cd}
\kappa_{cd}^{~~pq}{F}_{pq},
\eea
which differs only by a total minus sign from the main text.

\begin{widetext}
Then the Chern-Simons like term in differential form reads as
\bea&&
\hf\int \theta(x)F\wedge{F}=
\frac{1}{2^3}\int\theta(x)F_{ab}e^a\wedge{e^b}\wedge{F}_{cd}e^c\wedge{e^d}
\stackrel{!}{=}
\frac{1}{2}\int\sqrt{-g}d^4x\,{^*F}^{cd}\nabla_c\theta(x)A_d,
\eea
where $\stackrel{!}{=}$ means equal up to a boundary term
and we have used the Bianchi identity $\epsilon^{abcd}\nabla_cF_{ab}=0$ in the second to last equality.
\end{widetext}

\end{document}